\documentclass[aps,pra,showpacs,amssymb,nofootinbib,superscriptaddress,twocolumn]{revtex4-1}
\usepackage{tikz}
\usetikzlibrary{shapes,snakes,backgrounds,fit,decorations.pathreplacing}
\usepackage[ansinew]{inputenc}
\usepackage{bbm}
\usepackage{bm}
\usepackage{amsbsy}
\usepackage{amsthm}
\usepackage{amssymb}
\usepackage{amsfonts}
\usepackage{amsmath}
\usepackage{dsfont} 
\usepackage{graphicx} 
\usepackage{epsfig}
\usepackage{epstopdf}
\usepackage{dsfont}
\usepackage{color}
\usepackage{tabularx}
\usepackage{float}

\usepackage[colorlinks]{hyperref}
\makeatletter
\newcommand\org@hypertarget{}
\let\org@hypertarget\hypertarget
\renewcommand\hypertarget[2]{%
  \Hy@raisedlink{\org@hypertarget{#1}{}}#2%
  }
\makeatother
\usepackage[figure,table]{hypcap}
\usepackage[framemethod=tikz]{mdframed}
\definecolor{mycolor}{rgb}{0.122, 0.435, 0.698}
\newmdenv[innerlinewidth=0.5pt, roundcorner=4pt,linecolor=mycolor,innerleftmargin=6pt,
innerrightmargin=6pt,innertopmargin=6pt,innerbottommargin=6pt]{mybox}
\usepackage{paracol}
\usepackage{tcolorbox}
\tcbset{colback=mycolor!5,colframe=mycolor,
fonttitle=\bfseries, float=htb}
\newtcolorbox[blend into=figures]{boxfigure}[3][]
{ float*=ht,width=\textwidth,lower separated=false, center upper,
title={#2},label= fig:#3,#1}
\newtcolorbox[blend into=figures]{smallboxfigure}[3][]
{float=t,lower separated=false, blend before title=colon hang,
title={#2}, label= fig:#3 ,#1}
\newtcolorbox{smallbox}[3][]
{float=t,lower separated=false, blend before title=colon hang,
title={#2}, label= fig:#3 ,#1}
\newtcolorbox[blend into=tables]{smallboxtable}[3][]
{float=t,lower separated=false, blend before title=colon hang,
title={#2}, label= table:#3 ,#1}
\newtcolorbox[blend into=tables]{bigboxtable}[3][]
{float*,lower separated=false, blend before title=colon hang, width = \textwidth,
title={#2}, label= table:#3 ,#1}
\newcolumntype{Z}{|>{\centering\arraybackslash}X}

\usepackage{MnSymbol}
\usepackage{enumerate}

\usepackage{mathtools}
\usepackage{cleveref}
\hypersetup{
	bookmarksnumbered,
	pdfstartview={FitH},
	citecolor={darkgreen},
	linkcolor={darkred},
	urlcolor={darkblue},
	pdfpagemode={UseOutlines}}
\definecolor{darkgreen}{RGB}{50,190,50}
\definecolor{darkblue}{RGB}{0,0,190}
\definecolor{darkred}{RGB}{238,0,0}
\usepackage{soul}
\newcommand{\pr}{^{\prime}}

\newcommand{\ket}[1]{\ensuremath{\left|\right.\!{#1}\!\left.\right\rangle}}

\newcommand{\bra}[1]{\ensuremath{\left\langle\right.\!{#1}\!\left.\right|}}

\newcommand{\ketbra}[2]{\ensuremath{|{#1}\rangle\langle{#2}|}}

\newcommand{\scpr}[2]{\ensuremath{\left\langle\right.\hspace*{-1pt} #1 \hspace*{-1pt}\left|\right.\hspace*{-1pt} #2 \hspace*{-1pt}\left.\right\rangle}}

\newcommand{\subtiny}[3]{\ensuremath{_{\hspace{#1 pt}\protect\raisebox{#2 pt}{\tiny{$ #3$}}}}}
\newcommand{\suptiny}[3]{\ensuremath{^{\hspace{#1 pt}\protect\raisebox{#2 pt}{\tiny{$ #3$}}}}}

\newcommand{\tr}{\textnormal{Tr}}
\newcommand{\djj}{d\kern-0.4em\char"16\kern-0.1em}

\newcolumntype{Y}{>{\centering\arraybackslash}X}

\makeatletter
\renewcommand{\p@subsection}{}
\renewcommand{\p@subsubsection}{}
\makeatother

\newcommand{\iqoqi}{Institute for Quantum Optics and Quantum Information (IQOQI), Austrian Academy of Sciences, Boltzmanngasse 3, A-1090 Vienna, Austria}
\newcommand{\lugano}{Faculty of Informatics, Universit{\`a} della Svizzera italiana, Via G. Buffi 13, 6900 Lugano, Switzerland}
\newcommand{\marseille}{Universit{\'e} d'Aix-Marseille, Centre de Saint-J{\'e}r{\^o}me, 13397 Marseille Cedex 20, France}
\newcommand{\brun}{Institute of Computer Science, Masaryk University, Botanick\'{a} 68a, 60200 Brno, Czech Republic}
\newcommand{\sas}{Institute of Physics, Slovak Academy of Sciences, D\'{u}bravsk\'{a} cesta 9, 845 11 Bratislava, Slovakia}
\newcommand{\uab}{Universitat Autonoma de Barcelona, 08193 Bellaterra, Barcelona, Spain}
\newcommand{\hwu}{Institute of Photonics and Quantum Sciences (IPaQS), Heriot-Watt University, Edinburgh, EH14 4AS, UK}

\begin{document}

\title{Measurements in two bases are sufficient for certifying high-dimensional entanglement}
\author{Jessica Bavaresco}
\affiliation{\iqoqi}
\author{Natalia Herrera Valencia}
\affiliation{\iqoqi}
\affiliation{\marseille}
\author{Claude Kl\"ockl}
\affiliation{\iqoqi}
\affiliation{\brun}
\author{Matej Pivoluska}
\affiliation{\iqoqi}
\affiliation{\brun}
\affiliation{\sas}
\author{Paul Erker}
\affiliation{\iqoqi}
\affiliation{\lugano}
\affiliation{\uab}
\author{Nicolai Friis}
\email{nicolai.friis@univie.ac.at}
\affiliation{\iqoqi}
\author{Mehul Malik}
\email{mehul.malik@univie.ac.at}
\affiliation{\iqoqi}
\affiliation{\hwu}
\author{Marcus Huber}
\email{marcus.huber@univie.ac.at}
\affiliation{\iqoqi}

\date{\today}

\begin{abstract}
High-dimensional encoding of quantum information provides a promising method of transcending current limitations in quantum communication. One of the central challenges in the pursuit of such an approach is the certification of high-dimensional entanglement. In particular, it is desirable to do so without resorting to inefficient full state tomography. Here, we show how carefully constructed measurements in two bases  (one of which is not orthonormal) can be used to faithfully and efficiently certify {bipartite} high-dimensional states and their entanglement for any physical platform. To showcase the practicality of this approach under realistic conditions, we put it to {the} test for photons entangled in their orbital angular momentum. In our experimental setup, we are able to verify  $9$-dimensional entanglement for {a pair of photons on a} $11$-dimensional subspace {each}, at present the highest amount certified without any assumptions on the {state.}
\end{abstract}

\maketitle


Quantum communication offers advantages such as enhanced security in quantum key distribution (QKD) protocols~\cite{LoKurtyTamaki2014} and increased channel capacities~\cite{BennettShorSmolinThapliyal2002} with respect to classical means of communication. All of these improvements, ranging from early proposals~\cite{BennettBrassardMermin1992} to recent exciting developments such as fully device-independent QKD~\cite{AcinBrunnerGisinMassarPironioScarani2007, VaziraniVidick2014}, rely on one fundamental phenomenon: quantum entanglement. Currently, the workhorse of most implementations is entanglement between qubits, i.e., between two-dimensional quantum systems (e.g., photon polarization). However, it has long been known that higher-dimensional entanglement can be useful in overcoming the limitations of qubit entanglement~\cite{CerfBourennaneKarlssonGisin2002, BarrettKentPironio2006}, offering better key rates~\cite{GroeblacherJenneweinVaziriWeihsZeilinger2006}, higher noise resistance~\cite{HorodeckiRPMK2007, EltschkaSiewert2014} and improved security against different attacks~\cite{HuberPawlowski2013}.

Attempting to capitalize on this insight, recent experiments have successfully generated and certified high-dimensional entanglement in different degrees of freedom. In particular, the canonical way of generating two-dimensional polarization entanglement in down-conversion processes already offers the potential for exploring entanglement in higher dimensions. This can be achieved by exploiting spatial degrees of freedom~\cite{SchaeffPolsterHuberRamelowZeilinger2015, GutierrezEsparzaPimentaMarquesMatosoSperlingVogelPadua2014}, orbital angular momentum (OAM)~\cite{KrennMalikErhardZeilinger2017, VaziriWeihsZeilinger2002, KulkarniSahuMaganaBoydJha2017}, energy-time based encodings~\cite{RiedmattenMarcikicZbindenGisin2002, ThewAcinZbindenGisin2004, MartinGuerreiroTiranovDesignolleFroewisBrunnerHuberGisin2017, JhaMalikBoyd2008}, or combinations thereof in hyper-entangled quantum systems~\cite{SteinlechnerEckerFinkLiuBavarescoHuberScheidlUrsin2017, BarreiroLangfordPetersKwiat2005}. High-dimensional quantum systems have recently also been explored in matter-based systems such as Cesium atoms~\cite{Anderson:2015fu} and superconducting circuits~\cite{Kumar:2016cs}. Thus, high-dimensional quantum systems are not only of fundamental interest but are also becoming more readily available.

In this context, the certification and quantification of entanglement in many dimensions is a crucial challenge since full state tomography {(FST)} for bipartite systems of local dimension $d$ requires measurements in $(d+1)^2$ global product bases  {(i.e., tensor product bases for the global state)}~\cite{BertlmannKrammer2008}{, which quickly becomes impractical in high dimensions. Due to the complexity of realizing measurements in high-dimensional spaces, previous experiments that aimed to certify entanglement dimensionality (also known as Schmidt number) often had to resort to assumptions about the underlying quantum state $\rho$, including, amongst others, conservation of angular momentum}~\cite{KrennHuberFicklerLapkiewiczRamelowZeilinger2014}{, subtraction of accidentals}~\cite{ErhardMalikKrennZeilinger2017}{, perfect correlations in a desired basis}~\cite{DadaLeachBullerPadgettAndersson2011}{, or the assumption that the experimentally generated state is pure}~\cite{Kues:2017db}{. Although relying on such assumptions allows for a plausible quantification of entanglement dimensionality, it is not enough for unambiguous certification, which is desirable for secure quantum communication based on high-dimensional entanglement. The certification of the Schmidt number of a state is crucial for this task since a high-dimensional entangled state with low Schmidt number is LOCC equivalent to a low-dimensional entangled state.} Hence, unwieldy or inefficient entanglement estimation would strongly mitigate possible advantages of high-dimensional encoding. It is therefore of high significance to determine efficient and practical strategies for certifying high-dimensional states and quantifying their entanglement.

Here, we present an {efficient adaptive} method that is tailored to better harvest the information about entangled states generated in a given experiment, without the need for any assumptions about the {(generally mixed)} underlying state and requiring measurements in only two global product bases, regardless of the dimension of the state. {Our certification method can be implemented in any physical platform that is suitable for high-dimensional quantum information processing. For the purpose of assumption-free state certification, we certify the fidelity $F(\rho,\Phi)$ of the experimental state $\rho$ to a previously identified suitable target state $\ket{\Phi}$. We show that measurements in only two global product bases, $\{\ket{mn}\}_{m,n}$ and $\{\ket{\tilde{i}\tilde{j^*}}\}_{i,j}$, are sufficient to select $\ket{\Phi}$ and to bound the fidelity from below by a quantity $\tilde{F}(\rho,\Phi)\leq F(\rho,\Phi)$. For the purpose of assumption-free entanglement certification and quantification, we use our fidelity bound $\tilde{F}(\rho,\Phi)$ to certify the Schmidt number of the state.}

{One of the most surprising consequences of our results is the fact that \emph{all pure bipartite quantum states in any dimension can be faithfully certified by measurements in only two global product bases}. We prove this statement by deriving a tight lower bound to the fidelity with an appropriately chosen target state. All that is required for this certification is an educated guess of the corresponding Schmidt bases, which can be inferred from the physical setup at hand for all typical quantum optical platforms. The more accurate the identification of these bases, the higher the confidence in the certified state.}

For any identified target state $\ket{\Phi}$, the fidelity bound becomes exact when the setup indeed generates the pure state $\ket{\Phi}$ or the mixed state obtained by dephasing $\ket{\Phi}$. We demonstrate that this method can be generalized to measurements in multiple {global product bases}, yielding $\tilde{F}\suptiny{0}{0}{(M)}(\rho,\Phi)$, {in which $M+1$ is the total number of measurements bases, and in prime dimensions the fidelity bounds using measurements in $d+1$ bases $(M=d)$ become exact for all states, i.e., $\tilde{F}\suptiny{0}{0}{(d)}(\rho,\Phi)= F(\rho,\Phi)$.} Moreover, deriving general decompositions for dephased maximally entangled states further allows us to prove that unbiased measurement bases indeed provide a necessary and sufficient condition for tight \emph{Schmidt number bounds} for all pure states $\rho=\ket{\Phi}\!\!\bra{\Phi}$ and for maximally entangled states subject to pure dephasing. Our method can also be used for \emph{entanglement quantification} by providing lower bounds on the entanglement of formation~\cite{BennettDiVincenzoSmolinWootters1996,Wootters1998}. Here, our bounds outperform previous methods in terms of their noise robustness and the number of certified e-bits~\cite{ErkerKrennHuber2017}. {Finally, our bounds are also shown to be applicable for the certification of certain types of multipartite quantum states.}

To put these theoretical predictions to the test in realistic circumstances with actual noise, we devise and carry out an experiment based on photons entangled in their orbital angular momentum, allowing our approach to prove its mettle. In our experimental implementation, measurements are realised using computer programmable holograms implemented on spatial light modulators (SLMs). Employing the theoretical methods developed here, we are able to certify high target-state fidelities and verify record entanglement dimensionality: $9$-dimensional entanglement in $11$-dimensional subspaces, without any assumptions on the state itself. We use our experimental setup to fully explore the performance of our criteria for non-maximally entangled states, showcasing the flexibility of the derived results.
\\


\noindent\textbf{Entanglement dimensionality}
\\

Consider a typical laboratory situation for preparing a high-dimensional quantum system in a bipartite state $\rho$ that is to be employed for quantum information processing between two parties. In order to be useful, this state should be close to some highly entangled target state that is normally required to have a high purity. Let us therefore consider a pure target state $\ket{\Phi}$ with a desired Schmidt rank $k=k_{\mathrm{max}}$. The Schmidt rank is a measure of the entanglement dimensionality of the state and represents the minimum number of levels one needs to faithfully represent the state and its correlations in any {global product} basis. 
Ideally, the target state's Schmidt rank is equal (or close) to the (accessible) local dimension, $k_{\mathrm{max}}=d$, where we take the local Hilbert spaces to have the same dimension, {$\text{dim}(\mathcal{H}\subtiny{0}{0}{A})=\text{dim}(\mathcal{H}\subtiny{0}{0}{B})=d$}. For mixed states $\rho$ the Schmidt rank generalizes to the Schmidt number
\begin{align}
    k(\rho) &=\inf_{\mathcal{D}(\rho)} \left\{ \max_{\ket{\psi_{i}}\,\in\{(p_{i},\ket{\psi_{i}})\}_{i}} \Bigl\{ \operatorname{rank}\bigl(\tr\subtiny{0}{0}{B}\ket{\psi_{i}}\!\!\bra{\psi_{i}}\bigr)\Bigr\}\right\},
    \label{eq:Schmidt rank def}
\end{align}
where the infimum is taken over all pure state decompositions, i.e., $\mathcal{D}(\rho)$ is the set of all sets $\{(p_{i},\ket{\psi_{i}})\}_{i}$ for which $\rho=\sum_{i}p_{i}\ketbra{\psi_{i}}{\psi_{i}}$, $\sum_{i}p_{i}=1$, and $0\leq p_{i}\leq1$. 

The Schmidt number hence quantifies the maximal local dimension in which any of the pure state contributions to $\rho$ can be considered to be entangled and we hence call $k$ the \emph{entanglement dimensionality} of $\rho$. Note that this implies $k\leq d$. For example, any two-qubit entangled state (for which $d=2$) has an entanglement dimensionality $k=2$. A higher-dimensional entangled state, like a two-qutrit state ($d=3$), could have Schmidt number $k=3$, in which case it would indeed carry qutrit entanglement, or it could have only $k=2$, in which case the state would be LOCC equivalent to a two-qubit entangled state. In the latter example, even though the state has a higher local dimension, the entanglement dimensionality, which is our quantity of interest, is not higher. Trivially, all separable states have $k=1$.
\\

\noindent\textbf{Target state identification}
\\

The task at hand is to certify that the state $\rho$ generated in the lab is indeed close to the intended target state $\ket{\Phi}$ and thus provides the desired high-dimensional entanglement. One immediate first approach is to start with local projective measurements in the local Schmidt bases, i.e., the global product basis $\{\ket{mn}\}_{m,n=0,\ldots,d-1}$, which we designate as our \emph{standard basis}. These bases can typically be identified from conserved quantities or the setup design, but depending on the physical setup, the corresponding measurements are realised in different ways. In essence, a good {choice for the standard basis} provides a good target state. For instance, in an optical setting using OAM (as we employ in the experiment reported in this article) the chosen standard basis is the Laguerre-Gauss (LG) basis. In this case, these measurements are performed by coincidence post-selection after local {projective filtering}. That is, SLMs programmed with the phase pattern of a specific state $\ket{mn}$ act as local unitary operations, which are followed by single mode fibers (SMF) as local filters, and the number $N_{mn}$ of coincidences between local photon detectors is counted for each setting corresponding to fixed values of $m$ and $n$. In this way one can obtain the matrix elements
\vspace*{-3mm}
\begin{align}
    \bra{mn}\rho\ket{mn}    &=\frac{N_{mn}}{\sum_{k,l}N_{kl}}.
    \label{eq:matrix elements in terms of coincidence counts}
\end{align}

{A measurement in one global product basis can be realized by one $d$-outcome local measurement or equivalently replaced by $d$ single-outcome local measurements. The latter case employs the use of} $d$ local {filter settings} ($d^{2}$ {filter} settings globally) to obtain the values $\bra{mm}\rho\ket{mm}$. These are used to nominate a \emph{target state} $\ket{\Phi}=\sum_{m=0}^{d-1}\lambda_{m}\ket{mm}$ by identifying
\begin{align}
    \lambda_{m} &= \sqrt{\frac{\bra{mm}\rho\ket{mm}}{\sum_{n} \bra{nn}\rho\ket{nn}}}.
    \label{eq:lambdas}
\end{align}

This association alone by no means guarantees that the state $\rho$ really is equivalent to the target state $\ket{\Phi}$. Although the information about the diagonal elements of $\rho$ provides an informed guess, it is not enough to infer entanglement properties. In order to access this information, one could in principle perform costly {FST}. This requires {measurements in $(d+1)^{2}$ global product bases}~\cite{BertlmannKrammer2008}, {which is equivalent to $d^2(d+1)^{2}$ global filter settings.} Here, we propose a much more efficient alternative method to obtain a lower bound on the Schmidt rank of $\rho$ and on its fidelity to the target state. 
\\


\noindent\textbf{Dimensionality witnesses}
\\

For the certification of the Schmidt rank of $\rho$ we consider the fidelity $F(\rho,\Phi)$ to the target state $\ket{\Phi}$, given by
\begin{align}
    F(\rho,\Phi) &=\,\tr \bigl(\ket{\Phi}\!\!\bra{\Phi}\rho\bigr)
    =\sum_{m,n=0}^{d-1} \lambda_{m}\lambda_{n} \bra{mm} \rho \ket{nn}.
    \label{eq-fid2}
\end{align}
For any state $\rho$ of Schmidt rank $k\leq d$ the fidelity of Eq.~(\ref{eq-fid2}) is bounded by~\cite{PianiMora2007, FicklerLapkiewiczHuberLaveryPadgettZeilinger2014}
\begin{align}
    F(\rho,\Phi)   &\leq\,B_{k}(\Phi)\,:=\,\sum_{m=0}^{k-1}\lambda_{i_{m}}^{2},
    \label{eq:fid bound}
\end{align}
where the sum runs over the $k$ largest Schmidt coefficients, i.e., $i_{m}$, cyan{with} $m\in\{0,\ldots,d-1\}$ such that $\lambda_{i_{m}}\geq\lambda_{i_{m\pr}}\,\forall\,m\leq m\pr$. Consequently, any state for which $F(\rho,\Phi)>B_{k}(\Phi)$ is incompatible with a Schmidt rank of $k$ or less, implying an entanglement dimensionality of at least $k+1$. 
\\


\noindent\textbf{Fidelity bounds}
\\

The next step is hence to experimentally estimate the value of the fidelity $F(\rho,\Phi)$. To see how this can be done, we split the fidelity into two contributions, one that depends on the terms of Eq.~\eqref{eq-fid2} that are diagonal in the basis $\{\ket{mn}\}_{m,n}$, which will be called $F_{1}(\rho,\Phi)$, and the other that depends on the off-diagonal terms, called $F_{2}(\rho,\Phi)$ (see Methods).

The contribution $F_{1}(\rho,\Phi)$ can be calculated directly from the already performed measurement{s} in the basis $\{\ket{mn}\}_{m,n}$. However, exactly determining the term $F_{2}(\rho,\Phi)$ would require a number of measurements that scales with the dimension. To avoid such a high overhead, we employ bounds for $F_2(\rho,\Phi)$ that can be calculated from measurements in only one additional basis $\{\ket{\tilde{j}}\}_{j}$ (see Methods).

Using the previously obtained values $\{\lambda_{m}\}_{m}$, we define the basis $\{\ket{\tilde{j}}\}_{j=0,\ldots,d-1}$ according to
\begin{align}
    \ket{\tilde{j}} &= \frac{1}{\sqrt{\sum_{n}\lambda_{n}}} \sum_{m=0}^{d-1} \omega^{jm}\sqrt{\lambda_{m}} \ket{m},
\label{eq:tiltedbasis}
\end{align}
where $\omega=e^{2\pi i/d}$ and $\{\ket{m}\}_{m}$ is the standard basis. Notice that, although the basis vectors $\ket{\tilde{j}}$ are normalized by construction, they are not necessarily orthogonal, but become orthogonal and even mutually unbiased w.r.t. to $\{\ket{m}\}_{m}$ when all $\lambda_{m}$ are the same. We hence refer to $\{\ket{\tilde{j}}\}_{j}$ as the \textit{tilted basis}.

Due to this general non-orthogonality, the relation of Eq.~(\ref{eq:matrix elements in terms of coincidence counts}) between the diagonal matrix elements $\bra{\tilde{i}\tilde{j}^{*}}\rho\ket{\tilde{i}\tilde{j}^{*}}$ and the coincidence counts $\tilde{N}_{ij}$ for the local {filter} setting $\ket{\tilde{i}\tilde{j}^{*}}$ requires a small modification in terms of an additional normalization factor $c_{\lambda}:=\tfrac{d^{2}}{(\sum_{k}\lambda_{k})^{2}}\sum\limits_{m,n}\lambda_m\lambda_n\bra{mn}\rho\ket{mn}$, i.e.,
\begin{align}
    \bra{\tilde{i}\tilde{j}^{*}}\rho\ket{\tilde{i}\tilde{j}^{*}} &= \frac{\tilde{N}_{ij}}{\sum_{k,l}\tilde{N}_{kl}}\,c_{\lambda}.
    \label{eq:tilted basis matrix elements from coincidences}
\end{align}
Apart from the inclusion of $c_{\lambda}$ (see detailed derivation in the Supplementary Information), measurements in the tilted basis are in principle not different from measurements in any orthonormal basis.

The terms of Eq.~\eqref{eq:tilted basis matrix elements from coincidences}, along with the measurement results in the standard basis, allow us to bound the fidelity term $F_{2}(\rho,\Phi)$, which in turn provides a lower bound $\tilde{F}(\rho,\Phi)$ for the fidelity $F(\rho,\Phi)$ that is experimentally easily accessible.

We thus immediately obtain the dimensionality witness inequality
\begin{align}
    \tilde{F}(\rho,\Phi) &\leq F(\rho,\Phi)\leq B_{k}(\Phi),
    \label{eq:quitter dim witness}
\end{align}
which is satisfied by any state $\rho$ with Schmidt rank $k$ or less. Conversely, the entanglement dimensionality $d_{\mathrm{ent}}$ that is certifiable with our method is the maximal $k$ such that $\tilde{F}(\rho,\Phi)>B_{k-1}(\Phi)$.

A detailed derivation of this bound along with the proofs of its tightness can be found in the Methods section. In the Supplementary Information we further present a generalization of the fidelity bound to multiple measurement bases, the derivation of bounds for entanglement of formation that arise from our method, and an extension of our fidelity bound to a family of multipartite states.

Crucially, our witness requires only $2$ global product bases to be evaluated, and is hence significantly more efficient than the {$d+1$ and $(d+1)^{2}$ bases required for the exact evaluation of the fidelity or even a FST, respectively. For projective filtering the overall number of filter settings is obtained by multiplying the number of required bases by $d^{2}$.} A comprehensive comparison of required number of measurement settings is given in Table~\ref{table:ms}.

\begin{table}[ht!]
\begin{center}
{\renewcommand{\arraystretch}{1.5}
\begin{tabular}{| c || c | c | c |}
	\multicolumn{4}{c}{Number of measurements} \\
	\hline
	\begin{small} {Method} \end{small} 	& \, {FST} \, &  \,$F(\rho,\Phi)$ \,	& \, $\tilde{F}(\rho,\Phi)$ \, \\
	\hline
	\begin{small} {Global product bases} \end{small}  & \,  $(d+1)^2$   \, 	& $d+1$			& $2$  \\
	\hline
	\begin{small} {Local filter settings} \end{small}  	& \,  $(d+1)^2d^2$	\, & $(d+1)d^2$		& $2d^2$  \\
	\hline
\end{tabular}
}
\end{center}
\caption{The table shows the number of required measurements for optimal {full state tomography (FST)}, optimal fidelity measurement [$F(\rho,\Phi)$], and to calculate the fidelity bounds presented in this work [$\tilde{F}(\rho,\Phi)$]. The first line corresponds to the necessary number of measured global product bases (which can be realised with  {at most $d+1$-outcome} local measurements), and the second line, the necessary number of local filter settings (which can be realised with single-outcome local measurements).}
\label{table:ms}
\end{table}


\noindent\textbf{Experimental certification of high-dimensional entanglement}
\\

\begin{figure}
	\includegraphics[width=0.48\textwidth,trim={0cm 0mm 0cm 0mm}]{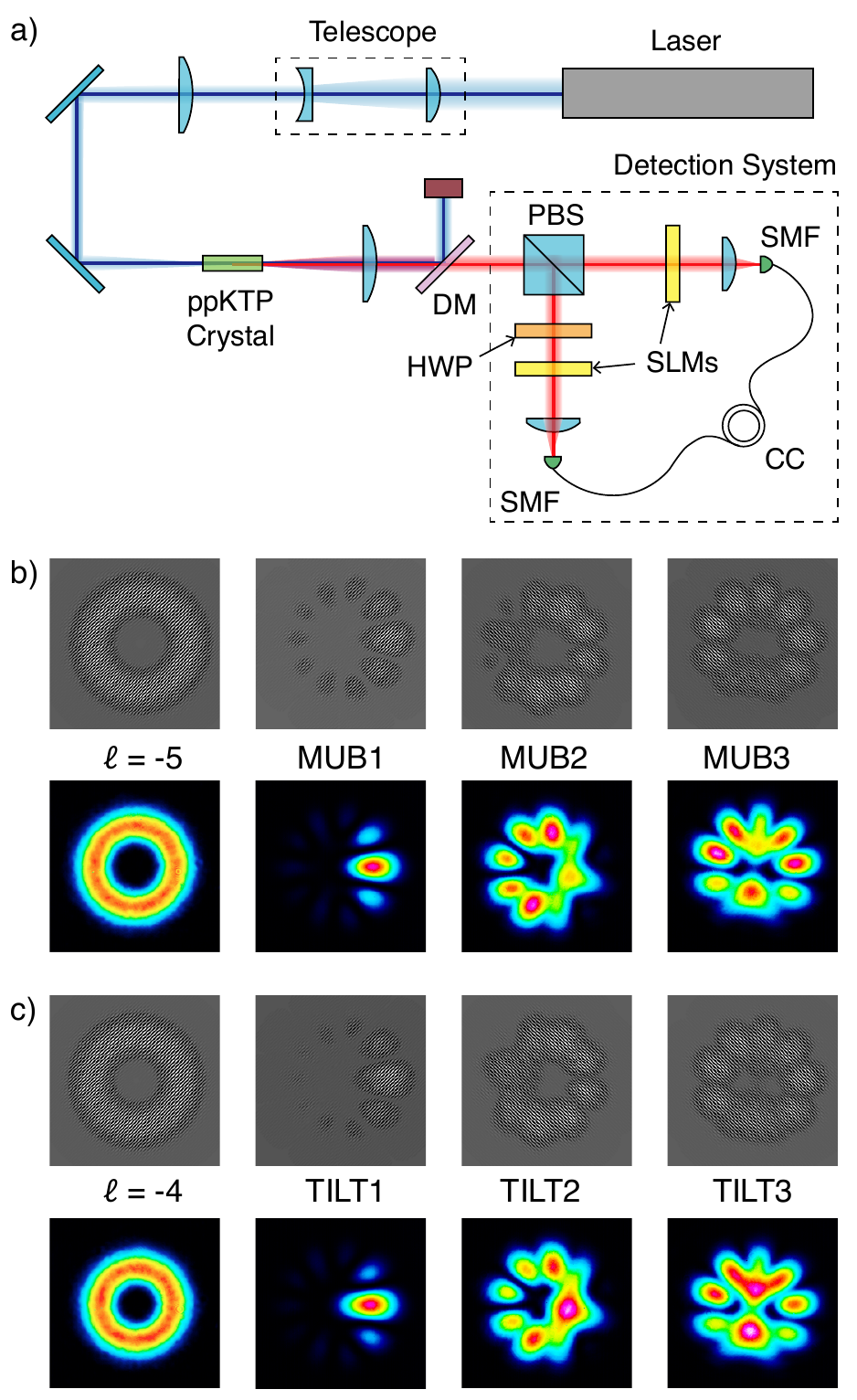}
\caption{\textbf{Experimental setup.} (a) A 405nm CW laser pumps a 5mm ppKTP crystal to generate a pair of infrared (810nm) photons via the process of Type-II spontaneous parametric down-conversion (SPDC), which are entangled in their orbital angular momentum (OAM). The pump is removed by a dichroic mirror (DM) and the two photons are separated by a polarizing beam splitter (PBS) and incident on two phase-only spatial light modulators (SLMs). A half-wave plate (HWP) is used to rotate the polarization of the reflected photon from vertical to horizontal, allowing it to be manipulated by the SLM. In combination with single-mode fibers (SMFs), the SLMs act as spatial mode filters. The filtered photons are detected by single-photon avalanche photodiodes (not shown) and time-coincident events are registered by a coincidence counting logic (CC);
 {(b) and (c) upper rows: examples of computer-generated holograms displayed on the SLMs for measuring the photons in a $d=11$ dimensional space; (b) and (c) left panels: standard LG basis modes with azimuthal quantum number $\ell=-5$ and $-4$; right panels of (b): 3 basis states from a MUB (denoted MUB1, MUB2, MUB3); right panel of (c): 3 basis states from a tilted basis [Eq.~}\eqref{eq:tiltedbasis} ](denoted TILT1, TILT2, TILT3); (b) and (c) lower rows: intensity images of the modes filtered by these holograms (see the Supplementary Information for details on how these intensity images were obtained).}
\label{fig:expsetup}
\end{figure}

We now apply our witness to certify high-dimensional orbital-angular-momentum (OAM) entanglement between two photons generated by Type-II SPDC in a non-linear ppKTP crystal (see Fig.~\ref{fig:expsetup}~(a) for details).
To this end, we display computer-programmed holograms [Fig.~\ref{fig:expsetup}~(b) {and (c)}] on spatial light modulators (SLMs) designed to manipulate the phase and amplitude of incident photons~\cite{ArrizonRuizCarradaGonzalez2007}. In this manner, we are able to projectively measure the photons in any spatial mode basis, e.g., the Laguerre-Gaussian (LG) basis, any mutually unbiased (MUB)~\cite{WoottersFields1989} or any tilted basis (TILT) composed of superpositions of elements of the standard basis [Eq.~(\ref{eq:tiltedbasis})].
Additional details of the experimental implementation, including information on the holograms, can be found in the Methods and Supplementary Information.

\begin{figure*}
	\includegraphics[width=0.98\textwidth,trim={0cm 0mm 0cm 0mm}]{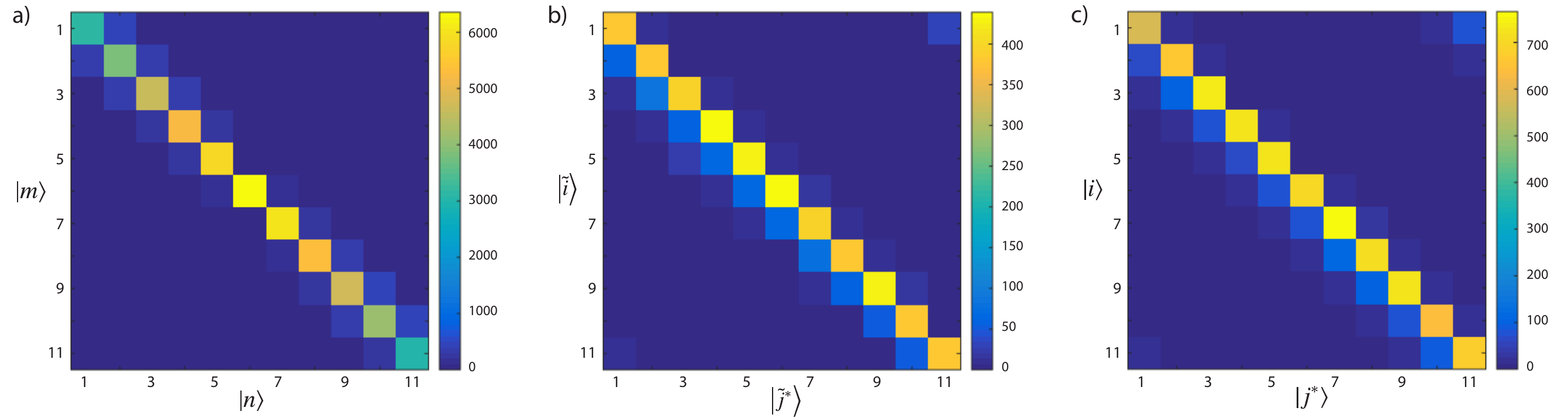}
\caption{\textbf{Experimental data certifying 9-dimensional entanglement.} Two-photon coincidence counts showing orbital angular momentum correlations in: (a) the standard LG basis $\{\ket{m},\ket{n}\}_{m,n}$, (b) the tilted basis $\{\ket{\tilde{i}},\ket{\tilde{j}^{*}}\}_{i,j}$, and (c) the first mutually unbiased basis $\{\ket{i},\ket{j^{*}}\}_{i,j}$. As seen in (a), our generated state is not maximally entangled (measured Schmidt coefficients $\lambda_m$ can be found in the Supplementary Information). For each set of two-basis measurements, we calculate a fidelity to the $d=11$ target state of $\tilde{F}(\rho,\Phi)= 76.2\pm0.6\%$ (LG and tilted bases), and $\tilde{F}(\rho,\Phi^{+})=74.8\pm0.4\%$ (LG and MUB). Even though the fidelity bound in the tilted case (b) is higher, the Schmidt number bounds are also higher and more difficult to overcome, yielding a certified entanglement dimensionality of $d_{\mathrm{ent}}=8$, slightly lower than the bound of $d_{\mathrm{ent}}=9$ obtained in the MUB case (c).}
\label{fig:dimthirteendata}
\end{figure*}

\begin{table}
\begin{center}
{\renewcommand{\arraystretch}{1.5}
\begin{tabular}{| c || c | c | c |}
	\multicolumn{4}{c}{Experimental results} \\
	\hline
	\ $d$ \   & \ $d_{\mathrm{ent}}$\ & \ $\tilde{F}(\rho,\Phi^{+})$\  & \ $\tilde{F}(\rho,\Phi)$\  \\
	\hline
	3   &   3        & 91.5$\pm$0.4\%  & 92.5$\pm$0.4\%  \\
	5   &   5        &  89.9$\pm$0.4\% &  90.0$\pm$0.5\% \\
	7   &   6        &  84.2$\pm$0.5\% & 86.9$\pm$0.6\%  \\
	11  &   9        &  74.8$\pm$0.4\% & 76.2 $\pm$0.6\% \\
\hline
\end{tabular}
}
\end{center}
\caption{Fidelities $\tilde{F}(\rho,\Phi^{+})$ and $\tilde{F}(\rho,\Phi)$ to the maximally entangled state and to the target state, obtained via measurements in two MUBs and two ($M=1$) tilted bases in dimension $d$, respectively. The second column lists the entanglement dimensionality $d_{\mathrm{ent}}$ certified using $\tilde{F}_{2}(\rho,\Phi^{+})$.}
\label{table:results}
\end{table}

For local dimensions up to $d=11$ (i.e., for azimuthal quantum numbers $\ell\in\{-5,\ldots,5\}$) we then proceed in the following way. First, we measure the two-photon state in the LG basis $\{\ket{m}\}_{m}$ to obtain a cross-talk matrix of coincidence counts $N_{mn}$ [Fig.~\ref{fig:dimthirteendata}~(a)], taking into account the effects of mode-dependent loss (see Supplementary Information). This allows us to calculate the density matrix elements $\bra{mn}\rho\ket{mn}$, estimate the $\lambda_m$, and nominate the target state $\ket{\Phi}$. We then use the set $\{\lambda_m\}_{m}$ to construct the \emph{tilted} basis $\{\ket{\tilde{j}}\}_{j}$ according to Eq.~(\ref{eq:tiltedbasis}) and perform correlation measurements [Fig.~\ref{fig:dimthirteendata}~(b)] that allow us to calculate $\bra{\tilde{j}\tilde{j}^{*}}\rho\ket{\tilde{j}\tilde{j}^{*}}$. From these measurements, we calculate the lower bound of the fidelity to the target state, for which we find high values, e.g., $\tilde{F}(\rho,\Phi)=76.2\pm0.6\%$ for $d=11$ (data for other dimensions is presented in Table~\ref{table:results}). However, in our setup, the certification thresholds $B_{k}$ for the tilted basis are higher than for the MUB (e.g., $B_{7}=0.72$ vs $B_{7}=0.64$ for $d=11$ in tilted versus MUB respectively). We therefore also measure the correlations in the first MUB $\{\ket{j}\}_{j}$ [Fig.~\ref{fig:dimthirteendata}~(c)] following the standard MUB construction by Wootters et al.~\cite{WoottersFields1989}, corresponding to $\lambda_m=1/\sqrt{d}$ for all $m$ in Eq.~(\ref{eq:tiltedbasis}). Using these measurements, we calculate lower bounds of the fidelity to the maximally entangled state, and find $\tilde{F}(\rho,\Phi^{+})=74.8\pm0.4\%$ for $d=11$, which is significantly above the bound of $B_{8}(\Phi^{+})=\tfrac{8}{11}\approx0.727$, but below $B_{9}(\Phi^{+})=\tfrac{9}{11}\approx0.818$. We hence certify 9-dimensional entanglement in this way. Note that the asymmetry in the counts just below and above the diagonal in Figs.~\ref{fig:dimthirteendata}~(b) and~(c) corresponds to a slight misalignment in the experiment. Errors in the fidelity are calculated by propagating statistical Poissonian errors in photon-count rates via Monte-Carlo simulation of the experiment. This demonstrates that our witness indeed works for efficiently certifying high-dimensional entanglement. Moreover, this shows that although the tilted basis measurements can achieve higher fidelities, one pays a price in terms of increased certification thresholds, and thus an increased sensitivity to noise.

Our approach hence provides a lower bound for $F(\rho,\Phi)$ and $k(\rho)$ using measurements in as little as $2$ global product bases. Each of these are realized by $d$ local filter settings on each side, totalling to $2d^2$ global filter settings instead of $d^{2}(d+1)^{2}$ for FST. For our state in a $11\times 11$-dimensional Hilbert space this corresponds to 242 filter settings, versus the 17,424 filter settings required for FST, which is a reduction by two orders of magnitude.\\


\noindent\textbf{Discussion and outlook}\\

A remarkable trait of high-dimensional entanglement is that measurements in two bases are enough to certify any entangled pure state for arbitrarily large Hilbert space dimension. We make use of this insight to establish fidelity bounds for states produced under realistic laboratory conditions. Using two (or, if desirable more, see Supplementary Information) local basis choices, these bounds can be employed to certify the Schmidt rank and entanglement of formation in a much more efficient way than is possible via full state tomography or even complete measurements of the fidelity. It is also interesting to note that the two measurement bases required for optimal fidelity certification become unbiased whenever the target state is maximally entangled. This procedure could be viewed as a trusted device analogue to self-testing~\cite{ColadangeloGohScarani2017}, requiring significantly fewer measurements and exhibiting a much greater noise resistance.

The strength of our method has its origin in the fact that we use readily available knowledge about the quantum system under investigation in terms of an educated guess for the Schmidt bases. This is close in spirit to assumptions commonly used in many experiments where preserved quantities in non-linear processes are harnessed to create entanglement. For the case of our experimental setup, this amounted to the conservation of transverse momenta~\cite{VaziriWeihsZeilinger2002}. Using holograms and couplings to single-mode fibers essentially implements single-outcome measurements (projective filtering), leading to $2d^2$ {filter settings globally}. This could be further improved by means of a mode sorter~\cite{BerkhoutLaveryCourtialBeijersbergenPadgett2010,Mirhosseini:2013em}, reducing the global measurement settings to merely $2$ (see Table~\ref{table:ms} for a comprehensive overview) {at the cost of using $d$-coincidence detectors}. But our proposed method is not limited to transverse momenta and OAM. In energy-time based setups~\cite{RiedmattenMarcikicZbindenGisin2002}, conservation of energy leads to the frequency or time-bin basis to be the natural Schmidt basis. Canonically these systems even feature $d$-outcome measurements, making them ideal candidates for {the} application of our method. Indeed, the states generated in the time-bin basis are generically close to being maximally entangled~\cite{MartinGuerreiroTiranovDesignolleFroewisBrunnerHuberGisin2017} and thus the tilted measurement would ideally be close to {MUBs}. There are various proposals as to how mutually unbiased measurements could also be directly implemented as $d$-outcome measurements in such systems~\cite{BroughamBarnett2013, Moweretal2013}. Finally, our method can be directly implemented using multi-path interferometers~\cite{SchaeffPolsterHuberRamelowZeilinger2015} where the natural Schmidt basis is the path degree of freedom. Let us stress again, however, that even deviations from the assumed situation do not invalidate the bounds employed in our approach, but lead (at most) to suboptimal performance, and an unambiguous certification is still ensured.

To demonstrate the practical utility of our method, we have performed an experiment using two photons entangled in their orbital angular momenta. We were able to certify  {$9$-dimensional} entanglement in a $11\times11$-dimensional Hilbert space, which is the highest number achieved so far without further assumptions on the underlying quantum state. This is achieved using only two local, unbiased measurement{ bases} ($11$-outcomes each), which are realized by $242$ local filters and coincidence counting. Using similar measurements in the tilted bases we are able to achieve target state  {fidelities of $92.5\%$ in $3$ dimensions and $76.2\%$ in $11$ dimensions}.
As we have shown, the certification method proposed here is thus surprisingly robust to noise and enables straightforward and assumption-free entanglement characterization in realistic quantum optics experiments. This further illustrates the usefulness of MUBs for the detection of entanglement~\cite{SpenglerHuberBrierleyAdaktylosHiesmayr2012, GiovanniniRomeroLeachDudleyForbesPadgett2013, TascaRudnickiAspdenPadgettSoutoRibeiroWalborn2015,PaulTascaRudnickiWalborn2016,ErkerKrennHuber2017,TascaSanchezWalbornRudnicki2017,SchneelochHowland2017} and correlations~\cite{SauerweinMacchiavelloMacconeKraus2017}.

Our certification method can also be generalized to operate with more than two bases, enabling an adaptable increase in noise resistance when required, as discussed in the Supplementary Information.
There we also show how our bounds can be extended to certify entanglement of formation. Remarkably, this approach can also be generalized to Greenberger-Horne-Zeilinger (GHZ)-like multipartite states recently created using OAM~\cite{MalikErhardHuberKrennFicklerZeilinger2016,ErhardMalikKrennZeilinger2017}, making large multipartite states generated by the methods of Ref.~\cite{KrennMalikFicklerLapkiewiczZeilinger2016} certifiable in a scalable manner. We give a brief exposition of this result in the Supplementary Information.\\


\noindent\textbf{Acknowledgments}
\\

We want to thank Anton Zeilinger for many fruitful discussions and guidance regarding the experimental setup. We acknowledge funding from the Austrian Science Fund (FWF) through the START project Y879-N27 and the joint Czech-Austrian project MultiQUEST (I 3053-N27 and GF17-33780L). JB and MH acknowledge support from the ESQ Discovery Grant of the Austrian Academy of Sciences (\"{O}AW) project OESQ0002X2. PE acknowledges funding by the European Commission (STREP RAQUEL) and the Swiss National Science Foundation (SNF). MM acknowledges support from the QuantERA ERA-NET Co-fund (FWF Project I3553-N36).
\\









\newpage
\newpage

\noindent\textbf{METHODS}
\\

\noindent\textbf{Derivation of the fidelity lower bound}
\\

In this section, we provide a proof for the fidelity bound
\begin{align}
    F(\rho,\Phi) &\geq\tilde{F}(\rho,\Phi),
    \label{eq:quitter dim witness appendix}
\end{align}
i.e., the right-hand side of Eq.~(\ref{eq:quitter dim witness}) of the main text, where $F(\rho,\Phi)=F_{1}(\rho,\Phi)+F_{2}(\rho,\Phi)$ and $\tilde{F}(\rho,\Phi)=F_{1}(\rho,\Phi)+\tilde{F}_{2}(\rho,\Phi)$, each split into two contributions. Since the first of these, given by
\begin{align}
    F_{1}(\rho,\Phi)    &:=\sum_{m} \lambda_{m}^{2}\bra{mm}\rho\ket{mm},
\end{align}
is the same for both $F$ and $\tilde{F}$, we hence want to concentrate on showing that $F_{2}\geq\tilde{F}_{2}$, where
\begin{align}
    F_2(\rho,\Phi)   &:=\sum_{m\neq n} \lambda_{m}\lambda_{n}\bra{mm}\rho\ket{nn},
    \label{eq:F2 appendix}
\end{align}
whereas the lower bound to $F_2(\rho,\Phi)$ is
\begin{align}
    \tilde{F}_{2}    &:=\tfrac{\left(\sum_m\lambda_m\right)^2}{d}\sum\limits_{j=0}^{d-1}\bra{\tilde{j}\tilde{j}^{*}}\rho\ket{\tilde{j}\tilde{j}^{*}}
        -\!\!\!\sum_{m,n=0}^{d-1}\!\! \lambda_m\lambda_n \bra{mn}\rho\ket{mn}  \nonumber\\
	&-\hspace*{-4.5mm}
    \sum\limits_{\substack{m\neq m\pr\!,  m\neq n \\ n\neq n\pr\!, n\pr\neq m\pr\\}}\hspace*{-4.5mm}
    \tilde{\gamma}_{mm\pr nn\pr}\,
    \sqrt{\bra{m\pr n\pr}\rho\ket{m\pr n\pr}\bra{mn}\rho\ket{mn}},
    \label{eq:F2 bound tilted basis appendix}
\end{align}
{where the asterisk denotes complex conjugation of the vector components w.r.t. $\{\ket{m}\}_{m}$} and the prefactor $\tilde{\gamma}_{mm\pr nn\pr}$ is given by
\begin{align}
    \tilde{\gamma}_{mm\pr nn\pr} &=
    \begin{cases}0\ \ \mbox{if}\ \ (m-m\pr-n+n\pr)\!\!\!\!\mod d \neq0\\[1mm]
    \displaystyle
    \sqrt{\lambda_m\lambda_n\lambda_{m\pr}\lambda_{n\pr}}\ \ \mbox{otherwise},
    \end{cases}
    \label{eq:tilde gamma appendix}
\end{align}
as we will show in the following. Here, the quantity $F_{1}(\rho,\Phi)$, as well as the second and third terms of $\tilde{F}_{2}$ in Eq.~(\ref{eq:F2 bound tilted basis appendix}) can be obtained directly from measurements in the standard basis $\{\ket{mn}\}_{m,n}$, whereas the first term of $\tilde{F}_{2}$ is constructed from diagonal density matrix elements w.r.t. to the tilted bases with elements
\begin{align}
    \ket{\tilde{j}} &= \frac{1}{\sqrt{\sum_{n}\lambda_{n}}} \sum_{m=0}^{d-1} \omega^{jm}\sqrt{\lambda_{m}} \ket{m},
    \label{eq:tilted basis appendix}
\end{align}
where $\omega=e^{2\pi i/d}$.
This non-orthogonal construction is motivated by the observations that $\ket{\Phi}$ is in general non-maximally entangled and that the tilted basis interpolates between the measurement bases required to obtain unit fidelities for pure product states $\ket{\Phi}=\ket{mn}$ (where the standard basis suffices) and for maximally entangled states $\ket{\Phi}=\ket{\Phi^{+}}$ (where the tilted basis becomes unbiased w.r.t. to the standard basis). The tilted basis $\{\ket{\tilde{j}}\}_{j}$ can be seen as a particular construction of a basis that satisfies the condition $|\scpr{m}{\tilde{j}}|^{2}=\lambda_m\lambda_j \forall m,j$ with the standard basis $\{\ket{m}\}_{m}$. Notice that the standard definition of mutually unbiased bases (MUBs) is recovered when $\lambda_i=\frac{1}{\sqrt{d}} \forall i$.

For the proof, we then focus on the matrix elements obtained from measurements w.r.t. the tilted basis. That is, we define the quantity
\begin{align}
    \Sigma  &:=\sum\limits_{j=0}^{d-1}\bra{\tilde{j}\tilde{j}^{*}}\rho\ket{\tilde{j}\tilde{j}^{*}}
    =\,\tfrac{1}{(\sum_{k}\lambda_{k})^{2}}
    \sum\limits_{\substack{m,m\pr \\ n,n\pr}}
    \sqrt{\lambda_{m}\lambda_{n}\lambda_{m\pr}\lambda_{n\pr}}
    \nonumber\\[1mm]
    &\hspace*{15mm}\times\sum\limits_{j=0}^{d-1}\omega^{j(m-m\pr-n+n\pr)}
    \bra{m\pr n\pr}\rho\ket{mn}.\label{eq:Sigma}
\end{align}
The sums over the standard basis components can then be split into several contributions. When $m=m\pr$ and $n=n\pr$, the phases all cancel, the sum over the tilted basis elements has $d$ equal contributions, and we hence have
\begin{align}
    \Sigma_{1}  &:=\,
    \tfrac{d}{(\sum_{k}\lambda_{k})^{2}}\sum\limits_{m,n}\lambda_{m}\lambda_{n}\bra{mn}\rho\ket{mn}.
\end{align}
When $m=m\pr$ but $n\neq n\pr$ (or vice versa) one finds terms containing the sum
\begin{align}
    \sum\limits_{j=0}^{d-1} \omega^{j(n\pr-n)}  &=\delta_{nn\pr}.
    \label{eq:orthogonality of standard basis}
\end{align}
Since $n\neq n\pr$, these terms vanish. For all remaining contributions to $\Sigma$ one has $m\neq m\pr$ and $n\neq n\pr$. These terms then again split into three sets. First, for $m=n$ and $m\pr=n\pr$ we recover the desired terms of the form
\begin{align}
    \Sigma_{2}  &:=\,\tfrac{d}{(\sum_{k}\lambda_{k})^{2}}\sum\limits_{m\neq n}\lambda_{m}\lambda_{n}\bra{mm}\rho\ket{nn},
\end{align}
which also appear in $F_2(\rho,\Phi)$ in Eq.~(\ref{eq:F2 appendix}). The terms where $m=n$ but $m\pr\neq n\pr$ (or vice versa) again vanish due to Eq.~(\ref{eq:orthogonality of standard basis}). Finally, this leaves the term
\begin{align}
    \Sigma_{3}  &:=
    \tfrac{1}{(\sum_{k}\lambda_{k})^{2}}
    \sum\limits_{\substack{m\neq m\pr \\ m\neq n \\ n\neq n\pr \\ n\pr\neq m\pr}}
    \sqrt{\lambda_m\lambda_n\lambda_{m\pr}\lambda_{n\pr}}
    \label{eq:sigma3}\\[-8mm]
    & \hspace*{28mm}\times\sum\limits_{j=0}^{d-1}\omega^{j(m-m\pr-n+n\pr)}
    \bra{m\pr n\pr}\rho\ket{mn},\nonumber\\[1mm]
    &=\,
    \tfrac{1}{(\sum_{k}\lambda_{k})^{2}}
    \sum\limits_{\substack{m\neq m\pr \\ m\neq n \\ n\neq n\pr \\ n\pr\neq m\pr}}
    \sqrt{\lambda_m\lambda_n\lambda_{m\pr}\lambda_{n\pr}}
    \nonumber\\[-8mm]
    & \hspace*{28mm}\times\operatorname{Re}\Bigl(c_{mnm\pr n\pr}
    \bra{m\pr n\pr}\rho\ket{mn}\Bigr),\nonumber
\end{align}
where we have used the abbreviation $c_{mnm\pr n\pr}:=\sum_{j}\omega^{j(m-m\pr-n+n\pr)}$. In the last step we have replaced $c_{mnm\pr n\pr}$ by its real part, since for each combination of values for $m,n,m\pr,n\pr$ the sum contains a term where the pairs $(m,n)$ and $(m\pr,n\pr)$ are exchanged. Each term in the sum is hence paired with another term that is its complex conjugate, and the total sum is hence real.

While $\Sigma_{1}$ and $\Sigma_{2}$ are accessible via measurements in the standard basis, the off-diagonal matrix elements in $\Sigma_{3}$ cannot be obtained from measurements w.r.t. $\{\ket{mn}\}_{m,n}$. In order to provide a useful lower bound for $\Sigma$ we therefore have to provide a bound for $\Sigma_{3}$. To this end, we can bound the real part by the modulus, i.e.,
\begin{align}
    &\operatorname{Re}\Bigl(c_{mnm\pr n\pr}
    \bra{m\pr n\pr}\rho\ket{mn}\Bigr) \,\leq\, |c_{mnm\pr n\pr}
    \bra{m\pr n\pr}\rho\ket{mn}|\nonumber\\[1mm]
    &\ \ \ \ =\,|c_{mnm\pr n\pr}|\cdot|\!\bra{m\pr n\pr}\rho\ket{mn}\!|\,.
    \label{eq:bound Re c times matrix element}
\end{align}
We then use the Cauchy-Schwarz inequality to bound the second factor on the right-hand side of~(\ref{eq:bound Re c times matrix element}) by writing $\rho=\sum_{i}p_{i}\ket{\psi_{i}}\!\!\bra{\psi_{i}}$ such that
\begin{align}
    |\!\bra{m\pr n\pr}\rho\ket{mn}\!|
    &=|\sum\limits_{i}
    \sqrt{p_{i}}\scpr{m\pr n\pr}{\psi_{i}}
    \sqrt{p_{i}}\scpr{\psi_{i}}{mn}|\nonumber\\[1mm]
    &\leq
    \sqrt{\sum\limits_{i}p_{i}\scpr{m\pr n\pr}{\psi_{i}}\scpr{\psi_{i}}{m\pr n\pr}}\nonumber\\[1mm]
    &\times\sqrt{\sum\limits_{i}p_{i}\scpr{mn}{\psi_{i}}\scpr{\psi_{i}}{mn}}
    \nonumber\\[1mm]
    &=\sqrt{\bra{m\pr n\pr}\rho\ket{m\pr n\pr}\bra{mn}\rho\ket{mn}}.
\end{align}
{In Eq.}~\eqref{eq:bound Re c times matrix element}, {note that in the first factor, $|c_{mnm\pr n\pr}|$, the sum $\sum_{j}\omega^{j(m-m\pr-n+n\pr)}$ vanishes whenever $(m-m\pr-n+n\pr)\mod d\neq0$, and equals to $d$ otherwise.} Collecting $c_{mnm\pr n\pr}/d$ with $\sqrt{\lambda_m\lambda_n\lambda_{m\pr}\lambda_{n\pr}}$ into $\tilde{\gamma}_{mm\pr nn\pr}$ as defined in Eq.~(\ref{eq:tilde gamma appendix}), this allows us to bound the quantity $\Sigma_{3}$ according to
\begin{align}
   \Sigma_{3}  &\leq\tfrac{d}{(\sum_{k}\lambda_{k})^{2}}\hspace{-5mm}
    \sum\limits_{\substack{m\neq m\pr \\ m\neq n \\ n\neq n\pr \\ n\pr\neq m\pr\\
	m-m\pr-n+n\pr = 0}}\hspace{-5mm}
    \tilde{\gamma}_{mm\pr nn\pr}
    \sqrt{\bra{m\pr n\pr}\rho\ket{m\pr n\pr}\bra{mn}\rho\ket{mn}}.\nonumber\\[-6mm]
\end{align}
Collecting the different contributions to $\Sigma$ we thus have
\begin{align}
    &\Sigma  \,=\,\Sigma_{1}+\Sigma_{2}+\Sigma_{3}\,=\,\sum\limits_{j=0}^{d-1}\bra{\tilde{j}\tilde{j}^{*}}\rho\ket{\tilde{j}\tilde{j}^{*}}\\[1mm]
    &\leq
    \tfrac{d}{(\sum_{k}\lambda_{k})^{2}}
    \Bigl(\sum\limits_{m,n}\lambda_{m}\lambda_{n}\bra{mn}\rho\ket{mn}
    +\!\sum\limits_{m\neq n}\lambda_{m}\lambda_{n}\bra{mm}\rho\ket{nn}
    \nonumber\\[1mm]
    &\ +\sum\limits_{\substack{m\neq m\pr \\ m\neq n \\ n\neq n\pr \\ n\pr\neq m\pr\\
	m-m\pr-n+n\pr = 0}}
    \tilde{\gamma}_{mm\pr nn\pr}\,
    \sqrt{\bra{m\pr n\pr}\rho\ket{m\pr n\pr}\bra{mn}\rho\ket{mn}}\,
    \Bigr).
    \nonumber
\end{align}
Conversely, this means that the term $F_{2}$ can be bounded by
\begin{align}
    F_{2}   &=\sum\limits_{m\neq n}\lambda_{m}\lambda_{n}\bra{mm}\rho\ket{nn}
    \label{}\\[1mm]
    &\geq\tfrac{(\sum_{k}\lambda_{k})^{2}}{d}\sum\limits_{j=0}^{d-1}\bra{\tilde{j}\tilde{j}^{*}}\rho\ket{\tilde{j}\tilde{j}^{*}}
    -\sum\limits_{m,n}\lambda_{m}\lambda_{n}\bra{mn}\rho\ket{mn}\nonumber\\
    &\ -\hspace{-4mm}\sum\limits_{\substack{m\neq m\pr \\ m\neq n \\ n\neq n\pr \\ n\pr\neq m\pr\\
	m-m\pr-n+n\pr = 0}}\hspace{-4mm}
    \tilde{\gamma}_{mm\pr nn\pr}\,
    \sqrt{\bra{m\pr n\pr}\rho\ket{m\pr n\pr}\bra{mn}\rho\ket{mn}},
    \nonumber
\end{align}
as claimed for the quantity $\tilde{F}_{2}$ in Eq.~(\ref{eq:F2 bound tilted basis appendix}). The fidelity $F(\rho,\Phi)$ can hence be bounded by measurements in only two local bases, $\{\ket{m}\}_{m}$ and $\{\ket{\tilde{j}}\}_{j}$, for each party{, i.e., two global product bases $\{\ket{mn}\}_{m,n}$ and $\{\ket{\tilde{i}\tilde{j^{*}}}\}_{i,j}$}.
\\


\noindent\textbf{Tightness of the fidelity bound}
\\

In this section, we show that whenever the system state $\rho$ is either equal to the (pure) target state $\rho=\ket{\Phi}\!\!\bra{\Phi}$ or is a dephased maximally entangled state $\rho_{\mathrm{deph}}(p)=p\ket{\Phi^{+}}\!\!\bra{\Phi^{+}}+\tfrac{1-p}{d}\sum_{m}\ket{mm}\!\!\bra{mm}$, the Schmidt number witness $\tilde{F}(\rho,\Phi)> B_{k-1}(\Phi)$ is not only a sufficient, but also a necessary condition for $\ket{\Phi}$ or $\rho_{\mathrm{deph}}$ to have a Schmidt rank larger or equal than $k$. For the state $\ket{\Phi}$ this is obvious. Since the coefficients $\lambda_{m}$ are determined by measurements in {the} Schmidt basis of $\rho=\ket{\Phi}\!\!\bra{\Phi}$, the fidelity bound is tight, and we have $\tilde{F}(\rho,\Phi)=F(\rho,\Phi)=1$ and $B_{k}(\Phi)$ is equal to $1$ if and only if $k=d$.

For dephased maximally entangled states we proceed by showing that there exists a Schmidt-rank $k$ state $\rho_{\mathrm{deph}}(p=p_{k})$ such that $F(\rho_{\mathrm{deph}}(p_{k}),\Phi)=B_{k}(\Phi)$ for every $k$. To this end, first note that $\rho_{\mathrm{deph}}$ can be written as
\begin{align}
    \rho_{\mathrm{deph}}   &=\,p\ket{\Phi^{+}}\!\!\bra{\Phi^{+}}+\tfrac{1-p}{d}\sum_{m}\ket{mm}\!\!\bra{mm}\nonumber\\[1mm]
    & =\,\tfrac{1}{d}\sum\limits_{m}\ket{mm}\!\!\bra{mm}+\tfrac{p}{d}\sum\limits_{m\neq n}\ket{mm}\!\!\bra{nn}\,,
    \label{eq:dephased max ent state}
\end{align}
which implies that $\lambda_{m}=\tfrac{1}{\sqrt{d}}\,\forall\,m$. That is, the corresponding target state is $\ket{\Phi}=\ket{\Phi^{+}}$ and $B_{k}=\tfrac{k}{d}$. The relevant fidelity then evaluates to
\begin{align}
    F(\rho,\Phi)    &=\,F(\rho_{\mathrm{deph}},\Phi^{+})\,=\,\tfrac{1+p(d-1)}{d}\,,
\end{align}
and $F(\rho_{\mathrm{deph}},\Phi^{+})=B_{k}$ for $p=p_{k}=\tfrac{k-1}{d-1}$. All we need to do now is to show that $\rho_{\mathrm{deph}}(p_{k})$ has a Schmidt rank no larger than $k$. To see this, consider the family of maximally entangled states in dimension $k$, i.e.,
\begin{align}
    \ket{\Phi^{+}_{\alpha}} &:=\tfrac{1}{\sqrt{|\alpha|}}\sum\limits_{m\in\alpha}\ket{mm}\,,
\end{align}
where $\alpha\subset \{0,1,\ldots,d-1\}$ with cardinality $|\alpha|=k$. In dimension $d$, we can find $\tbinom{d}{k}$ such states and consider their incoherent mixture, i.e.,
\begin{align}
    \rho_{k}    &=\,\frac{1}{\tbinom{d}{k}}\sum\limits_{\alpha\,\text{s.t.}\,|\alpha|=k}\ket{\Phi^{+}_{\alpha}}\!\!\bra{\Phi^{+}_{\alpha}}.
\end{align}
Since each of the ${\Phi^{+}_{\alpha}}$ has Schmidt rank $k$, the convex sum $\rho_{k}$ cannot have a Schmidt rank larger than $k$. Since there are $\tbinom{d-1}{k-1}$ terms contributing to every nonzero diagonal matrix element, we have $\bra{mn}\rho_{k}\ket{mn}=\tfrac{1}{d}\delta_{mn}$. Similarly, every nonvanishing off-diagonal matrix element has $\tbinom{d-2}{k-2}$ contributions, and we hence have $\bra{mn}\rho_{k}\ket{ij}=\tfrac{k-1}{d(d-1)}\delta_{mn}\delta_{ij}$ for $m\neq i$. It is then easy to see that the fidelity with the maximally entangled state (in dimension $d$) is $F(\rho_{k},\Phi^{+})=\tfrac{k}{d}$. More specifically, comparison with Eq.~(\ref{eq:dephased max ent state}) reveals that $\rho_{\mathrm{deph}}=\rho_{k}$ for $p=p_{k}=\tfrac{k-1}{d-1}$. Since the Schmidt rank of $\rho_{k}$ is smaller or equal than $k$, we have hence shown that the Schmidt rank of the dephased maximally entangled state $\rho_{\mathrm{deph}}(p_{k})$ with $F(\rho_{\mathrm{deph}}(p_{k}),\Phi)=B_{k}$ is $k$ or less. Consequently, $F(\rho_{\mathrm{deph}},\Phi^{+})> B_{k-1}$ is a necessary and sufficient condition for $\rho_{\mathrm{deph}}$ to have Schmidt rank $k$.

Moreover, since the fidelity bound $\tilde{F}\leq F$ is tight for $\rho_{\mathrm{deph}}$ already for $M=1$ and the tilted basis is unbiased w.r.t. the standard basis for dephased maximally entangled states, we can conclude that measurements in two unbiased bases provide the necessary and sufficient condition $\tilde{F}(\rho_{\mathrm{deph}},\Phi^{+})> B_{k-1}$ for Schmidt rank $k$ for these states.
\\


\noindent\textbf{Role of the target state}
\\

The initial designation of the target state $\ket{\Phi}$, or rather its Schmidt basis, helps to suitably adapt the dimensionality witness to the experimental situation. Although identifying the Schmidt basis from the setup could in principle be seen as an assumption about the underlying state, choosing a basis that is far from the Schmidt basis doesn't invalidate our certification method. Since the latter is based on lower-bounding the fidelity to the target state, such a misidentification would simply result in a reduced performance by using lower bounds on the fidelity to a state that is far from the actual state.  An analysis of how our fidelity bounds are affected by a ``wrong'' choice of basis is provided in the Supplementary Information. In other words, a non-optimal guess can lead to what is called a type-II-error (i.e., a ``false negative"), but never to a type-I-error (i.e., a ``false positive"). This means that a suboptimal guess of the target state may lead to a less than optimal value for the certified fidelity and/or Schmidt number. The entanglement dimensionality (Schmidt number) certified by a wrong choice of basis may hence be lower than the actual entanglement dimensionality (Schmidt number) of the underlying state $\rho$, but never higher. In summary, it can be concluded that the performance of our method may depend on the expected target state, but the method does not require any assumptions about the true system state $\rho$.

While this certification method is thus independent of the specific circumstances in the laboratory, it can be noted that it works particularly well for certain types of states. For instance, whenever the target state matches the underlying state up to pure dephasing, i.e., when $\rho=p\ket{\Phi}\!\!\bra{\Phi}+\tfrac{1-p}{d}\sum_{m}\ket{mm}\!\!\bra{mm}$, the fidelity bound $\tilde{F}(\rho,\Phi)\leq F(\rho,\Phi)$ is tight, since the last term in Eq.~(\ref{eq:F2 bound tilted basis appendix}) vanishes in this case.
Moreover, whenever these states are pure ($p=1$) or dephased maximally entangled states (arbitrary $p$ but $\ket{\Phi}=\ket{\Phi^{+}}$) one can further show that the Schmidt number bound $F(\rho,\Phi)\leq B_{k}(\Phi)$ is also tight (see Supplementary Information for derivation), in which case we have $\tilde{F}(\rho,\Phi)=F(\rho,\Phi)=B_{d_\text{ent}}(\Phi)$.

In addition, it can sometimes be helpful to select a ``wrong" target state on purpose. For example, the maximally entangled state $\ket{\Phi^{+}}=\tfrac{1}{\sqrt{d}}\sum_{m}\ket{mm}$, i.e., a target state whose coefficients where chosen to be $\lambda_{m}=\tfrac{1}{\sqrt{d}}\,\forall\,m$, may at times offer a higher Schmidt number lower bound than a target state with coefficients $\lambda_m$ taken from the measurement results in the standard basis, even though the fidelity bound would be lower. In the case of the maximally entangled target state, the tilted basis becomes an orthonormal basis that is mutually unbiased w.r.t. to the standard basis and we have $B_{k}(\Phi^{+})=\tfrac{k}{d}$. Since this bound is lower than for general values of $\lambda_m$, it may be easier to overcome, particularly in the presence of noise, and hence yield a higher certified Schmidt number. Indeed, this is the case in our experimental realization (see Table~\ref{table:results} of the main text), where higher fidelity bounds are attained with the tilted basis but higher Schmidt number is obtained using MUBs. It is important to point out again, however, that regardless of the choice of target state, the certified fidelity and Schmidt number will always be correct and never over-estimated. In practice this means that a bad choice of basis may lead to a worse noise resistance and it may be harder to certify any entanglement, but when one manages to certify it, this result can be trusted.
 \\


\noindent\textbf{Experimental details}
\\

Finally, let us discuss the experimental implementation of our entanglement certification method in more detail. As shown in Fig.~\ref{fig:expsetup}~(a) of the main text, our source consists of a single-spatial mode, continuous wave 405nm diode laser (Toptica iBeam Smart 405 HP) with $\sim$140mW of power. The laser is demagnified with a 3:1 telescope system of lenses and focused by a 500mm lens to a spot size of 330$\mu$m (1/$e^2$ beam diameter) at the ppKTP crystal. The 5mm long ppKTP crystal is designed for degenerate Type-II spontaneous parametric downconversion (SPDC) from 405nm to 810nm at 25$^\circ$C, and is housed in a custom-built oven for this purpose. The SPDC process generates orthogonally polarized pairs of photons entangled in the Laguerre-Gaussian (LG) basis. The photon pairs are recollimated by a 200mm lens, separated by a polarizing beamsplitter (PBS), and incident on phase-only spatial light modulators (SLMs).

The SLMs (Holoeye PLUTO) have a parallel-aligned LCOS design with a dimension of 15.36mm$\times$8.64mm, resolution of 1920$\times$1080 pixels, reflectivity of approximately 60\%, and a diffraction efficiency of 80\% at 810nm. The photons are transformed and reflected by these SLMs (shown in transmission for simplicity) and coupled into single-mode fibers (SMFs) with a coupling efficiency of approx.~50\%. The SMFs carry the photons to single-photon avalanche detectors (not shown, Excelitas SPCM-AQRH-14-FC) with a detection efficiency of 60\% at 810nm. The detectors are connected to a custom-built coincidence counting logic (CC) with a coincidence-time window of 5ns.

The SLMs and SMFs together act as projective filters for the photon spatial modes. The SLMs are used to display a computer-generated hologram (CGH) that multiplies the incident photon amplitude by an arbitrary amplitude and phase. In this manner, photons in a particular spatial mode (Laguerre-Gaussian or superpositions thereof) are converted to a fundamental Gaussian mode, which then effectively couples to the SMF. The manipulation of both the phase and amplitude of a photon by means of a phase-only device such as an SLM requires the design of a class of phase-only CGHs that allow one to encode arbitrary scalar complex fields. Following the Type 3 method in Ref.~\cite{ArrizonRuizCarradaGonzalez2007}, our CGH encodes the modulation of a complex field given by $s(x,y)=A(x,y)\exp[i\phi(x,y)]$ into a phase-only function whose functional form depends explicitly on the amplitude and phase of the field $s(x,y)$. This allows arbitrary complex amplitudes to be generated/measured by a phase-only device, albeit at the expense of additional loss. Additionally, we divide the measurement amplitude $s(x,y)$ by an offset fundamental Gaussian amplitude in order to maximize its overlap with the SMF mode.

A two-photon count rate of approximately 23,000 pairs/sec (Gaussian modes) is measured at the detectors (with blazed gratings displayed on the SLMs), and singles rates of 160,000 and 173,000 counts/sec in the reflected and transmitted PBS arms respectively. The resulting coincidence-to-singles ratios are consistent with the losses described above in each arm. The lossy complex amplitude hologram described above further reduces the two-photon Gaussian-mode count rate to 668 pairs/sec. These holograms have a mode-dependent loss that varies for different incident modes. In the Supplementary Information, we discuss how the coincidence and singles rates allow us to account for this mode-dependent loss. As shown in Fig.~\ref{fig:dimthirteendata}~(a) of the main text, the resultant state measured by these holograms in the standard Laguerre-Gaussian basis is close to $\ket{\Phi}=\sum_{m=0}^{10}\lambda_{m}\ket{mm}$, with 89\% counts on the diagonal.
The individual $\lambda_{m}$ values are: $\lambda_0=0.255$, $\lambda_1=0.259$, $\lambda_2=0.292$, $\lambda_3=0.315$, $\lambda_4=0.335$, $\lambda_5=0.349$, $\lambda_6=0.339$, $\lambda_7=0.316$, $\lambda_8=0.305$, $\lambda_9=0.272$, and $\lambda_{10}=0.260$. Note that $m\in\{0,\ldots,10\}$ corresponds to Laguerre-Gaussian modes with an OAM of $\ell\in\{-5,\ldots,5\}$. The measured state is correlated in OAM, as the reflection at the PBS flips the sign of one photon from the initially OAM-anti-correlated state.

The probability that one CW pump photon downconverts into a pair of photons in our ppKTP crystal is $10^{-9}$. While this is two orders of magnitude higher than $\beta$-BBO, it is still quite low. The corresponding probability of two pairs being produced simultaneously is then significantly lower at $10^{-18}$ and can be 
neglected. The rate of accidental counts becomes a factor when the singles rates are high and the measurement integration time is long. For example, in the Gaussian (brightest) modes, there are 6675 pairs measured in 10 seconds. The total singles are 230438 and 249617, 
an accidental rate of $\approx$2.9/sec. Correcting for accidental coincidences in in this manner increases the measured fidelities of our states slightly.






\end{document}


\title{Supplementary Information to ``Measurements in two bases are sufficient for certifying high-dimensional entanglement''}
\author{Jessica Bavaresco}
\affiliation{\iqoqi}
\author{Natalia Herrera Valencia}
\affiliation{\iqoqi}
\affiliation{\marseille} 
\author{Claude Kl\"ockl}
\affiliation{\iqoqi}
\affiliation{\brun}
\author{Matej Pivoluska}
\affiliation{\iqoqi}
\affiliation{\brun}
\affiliation{\sas}
\author{Paul Erker}
\affiliation{\iqoqi}
\affiliation{\lugano}
\affiliation{\uab}
\author{Nicolai Friis}
\email{nicolai.friis@univie.ac.at}
\affiliation{\iqoqi}
\author{Mehul Malik}
\email{mehul.malik@univie.ac.at}
\affiliation{\iqoqi}
\affiliation{\hwu}
\author{Marcus Huber}
\email{marcus.huber@univie.ac.at}
\affiliation{\iqoqi}

\date{\today}


\maketitle

\hypertarget{sec:appendix}
\appendix
\renewcommand{\thesubsubsection}{S.\Roman{subsection}.\arabic{subsubsection}}
\renewcommand{\thesubsection}{S.\Roman{subsection}}
\renewcommand{\thesection}{S}
\setcounter{equation}{0}
\numberwithin{equation}{section}
\setcounter{figure}{0}
\renewcommand{\thefigure}{S.\arabic{figure}}

In this supplemental material, we provide detailed proofs and additional calculations illustrating the versatility of the results presented in the main text, as well as more information on the experimental implementation. To provide some context, let us compactly summarize the main results:

\begin{smallbox}{Summary of main results}{fidel}
\textbf{Fidelity bound}: \ \ {$\tilde{F}\suptiny{0}{0}{(M)}(\rho,\Phi)\leq F(\rho,\Phi)$} 
\begin{itemize}
\item{{Obtained from measurements in $M+1$ \emph{global product} bases};}
\item{{\emph{Exact} for dephased pure states with only \emph{two} bases ($M=1$)};}
\item{Free of assumptions about the state $\rho$;}
\item{\emph{Exact} in prime dimensions for $M=d$;}
\item{Also works for certain classes of multipartite entangled states;}
\end{itemize}
\textbf{Schmidt number witness}: $\tilde{F}(\rho,\Phi)\Rightarrow d_\text{ent}$
\begin{itemize}
\item{\emph{Exact} for all pure states;}
\item{\emph{Exact} for dephased max. entangled states;}
\end{itemize}
\textbf{Entanglement bound}: $\tilde{F}(\rho,\Phi^{+})\Rightarrow \mathcal{E}_{\mathrm{oF}}(\rho)$
\begin{itemize}
\item{\emph{Improvement} w.r.t. previous bounds~\cite{ErkerKrennHuber2017}.}
\end{itemize}
\end{smallbox}

The basis for these results are measurements in two $(M=1)$ [or more $(M>1)$] global product bases, one of which \textemdash\ the standard basis $\{\ket{mn}\}_{m,n}$ \textemdash\ provides initial data (a set of values $\{\lambda_{m}\}$) that is used to construct the other (``tilted") basis. To summarize this method:

\begin{smallbox}{Adaptive strategy for certifying\ \  entanglement dimensionality}{strategy}
\label{fig:Adaptive strategy for certifying entanglement}
\begin{enumerate}[\hspace*{-3mm}(1)]
            \item{Identify standard basis $\{\ket{mn}\}$ and measure coincidences $\{N_{mn}\}$ to obtain  $\{\bra{mn}\rho\ket{mn}\}$.}
            \item{Calculate $\{\lambda_{m}\}$ and nominate target state $\ket{\Phi}$.}
            \item{Construct tilted basis $\{\ket{\tilde{j}}\}$ and measure coincidences $\{\tilde{N}_{ij}\}$ to obtain $\{\bra{\tilde{j}\tilde{j}^{*}}\rho\ket{\tilde{j}\tilde{j}^{*}}\}$.}
            \item{Evaluate $\tilde{F}(\rho,\Phi)$ and $B_{k=1}(\Phi)$, $\ldots$, $B_{k=d-1}(\Phi)$. The certified entanglement dimensionality is $d_{\mathrm{ent}}=\max\{k\,|\,\tilde{F}(\rho,\Phi)>B_{k-1}(\Phi)\}$.}
        \end{enumerate}
\end{smallbox}

To be more precise, the (first) local tilted basis $\{\ket{\tilde{j}}\}_{j=0,\ldots,d-1}$ is constructed from the local standard basis $\{\ket{m}\}_{m=0,\ldots,d-1}$ according to
\begin{align}
    \ket{\tilde{j}} &= \tfrac{1}{\sqrt{\sum_{n}\lambda_{n}}} \sum_{m=0}^{d-1} \omega^{jm}\sqrt{\lambda_{m}} \ket{m}.
    \label{eq:tiltedbasis}
\end{align}
To obtain the values $\{\lambda_{m}\}$, we use the following method. As explained in the main text, local filters [e.g., an appropriately programmed spatial light modulator (SLM)] are employed to allow only systems in particular states to be detected. For a particular setting with fixed $m$ and $n$ corresponding to the global orthonormal basis $\{\ket{mn}\}_{m,n}$ one then counts the coincidences $N_{mn}$, which give an estimate of the diagonal density matrix elements of the underlying system state $\rho$ via
\begin{align}
    \bra{mn}\rho\ket{mn}    &=\frac{N_{mn}}{\sum_{i,j}N_{ij}}.
    \label{eq:matrix elements in terms of coincidence counts appendix}
\end{align}
These matrix elements in turn determine the values
\begin{align}
    \lambda_{m} &= \sqrt{\frac{\bra{mm}\rho\ket{mm}}{\sum_{n} \bra{nn}\rho\ket{nn}}},
    \label{eq:lambdas}
\end{align}
which can be interpreted as nominating a \emph{target state} $\ket{\Phi}=\sum_{m=0}^{d-1}\lambda_{m}\ket{mm}$. Measurements in the second (tilted) basis (and potential additional tilted bases) then allow to evaluate a lower bound $\tilde{F}(\rho,\Phi)$ for the fidelity $F(\rho,\Phi)\geq\tilde{F}(\rho,\Phi)$ to the target state, as well of a number of threshold values $B_{k=1}(\Phi)$, $\ldots$, $B_{k=d-1}(\Phi)$. A Schmidt-rank of $k$ is then certified if the fidelity bound $\tilde{F}(\rho,\Phi)$ surpasses the value $B_{k-1}(\Phi)$, given by
\begin{align}
    B_{k}(\Phi) &:=\,\sum_{m=0}^{k-1}\lambda_{i_{m}}^{2}.
    \label{eq:fid bound}
\end{align}


Additional information on various aspects of this method and its implementation are given in the following.
Section~\ref{sec:appendix normalization tilted basis} details how measurements in the tilted basis can be performed.
In Sec.~\ref{sec:appendix noise robustness}, the noise robustness of our approach is discussed for the important special case of maximally entangled target states subject to white noise. In Sec.~\ref{sec:appendix more than one tilted basis}, we discuss the generalization of the fidelity bounds to measurements in more than two bases. We continue by discussing some simple bounds for the entanglement of formation in Sec.~\ref{sec:appendix EoF bounds}, before showing the connection to the fidelity bounds to the maximally entangled state and discussing the robustness of these quantification techniques in comparison to previous methods in Sec.~\ref{sec:appendix robustness of EoF bounds}. We show how the method can naturally be extended to the multipartite case in Sec.~\ref{sec:appendix multipartite}. In Sec.~\ref{sec:wrong schmidt basis} we analyse the effects of a non-ideal choice of the standard basis, while Sec.~\ref{sec:appendix LG Basis} shows evidence for the mutual unbiasedness of the implemented measurement bases. In Sec.~\ref{sec:appendix Implementations of MUBs in other setups}, we show an experimental example of a second spatial mode basis and discuss how mutually unbiased measurements can be readily implemented in a wide range of high-dimensional quantum systems using current technology. Finally, in Sec.~\ref{sec:syserror}, we discuss two sources of systematic error introduced by our specific measurement devices---mode-dependent loss and imperfect hologram measurements.


\subsection{Normalization for measurements in the tilted bases}\label{sec:appendix normalization tilted basis}

We now discuss in more detail how the measurements in the bases $\{\ket{mn}\}_{m,n}$ and $\{\ket{\tilde{i}\tilde{j^{*}}}\}_{i,j}$ can be performed by means of a post-selection procedure that we refer to as projective filtering. As explained above, estimates of the diagonal matrix elements $\bra{mn}\rho\ket{mn}$ of $\rho$ w.r.t. the standard basis can be obtained from coincidence counting. For the standard basis, one finds $\sum_{m,n}\bra{mn}\rho\ket{mn}=1$ by construction, which is sensible, since this expression corresponds to $\tr(\rho)$ for an orthonormal basis. In other words, $\sum_{m,n}\ket{mn}\!\!\bra{mn}=\mathds{1}$ is a resolution of the identity.

The same cannot be said for the (generally non-orthogonal) basis $\{\ket{\tilde{j}}\}_{j}$. However, the projectors $\{\ketbra{\tilde{j}}{\tilde{j}}\}_j$ can be used to construct a valid non-projective $(d+1)$-outcome positive operator-valued measure (POVM). The first $d$ elements of this POVM correspond to projectors in the tilted basis {divided by a factor of $d$}, while the last POVM element is obtained by subtracting the sum of the aforementioned {elements} from the identity, which results in a positive semi-definite operator, that is, the set of POVM elements for a measurement in a tilted basis is {$\left\{\frac{1}{d}\{\ketbra{\tilde{j}}{\tilde{j}}\}_{j=0,\ldots,d-1},\mathds{1}-\frac{1}{d}\sum_{j=0}^{d-1}\ketbra{\tilde{j}}{\tilde{j}}\right\}$}. By construction this is a $(d+1)$-outcome measurement{. However, when measurements are performed using projective filtering, only $d$ filter settings, corresponding to the $d$ projectors $\ket{\tilde{j}}\!\!\bra{\tilde{j}}$, need to be performed if the measurement results of the standard basis are already available. To see this, note that projective filtering implies that instead of the probabilities $p_{j}=\bra{\tilde{j}}\rho\ket{\tilde{j}}$ and $\bar{p}=\tr\bigl((\mathds{1}-\sum_{j=0}^{d-1}\ketbra{\tilde{j}}{\tilde{j}})\rho\bigr)=1-\sum_{j=0}^{d-1}p_{j}$, one obtains only the count rates $N_{j}=N p_{j}$ and $\bar{N}=N\bar{p}$, where $N$ is the overall number of photons such that $N=\bar{N}+\sum_{j=0}^{d-1}N_{j}$. The $d$ values $N_{j}$ alone hence do not fully determine the desired values $p_{j}=N_{j}/N$, but the normalization factor $N$ can be determined from $\sum_{j=0}^{d-1}N_{j}$ together with the measurements already performed in the standard basis $\{\ket{m}\}_m$, which yield $\sum_{j=0}^{d-1}p_{j}=\tfrac{1}{N}\sum_{j=0}^{d-1}N_{j}$. For the two-party scenario with measurements w.r.t. the global product basis $\{\ket{\tilde{i}\tilde{j^{*}}}\}_{i,j}$, this sum of density matrix elements in the tilted basis is calculated as}
\begin{align}
    \sum\limits_{i,j}\bra{\tilde{i}\tilde{j^{*}}}\rho\ket{\tilde{i}\tilde{j^{*}}}   &=\,\tfrac{1}{(\sum_{k}\lambda_{k})^{2}}\sum\limits_{\substack{m,m\pr\\ n,n\pr}}
    \sqrt{\lambda_m\lambda_n\lambda_{m\pr}\lambda_{n\pr}}\nonumber\\[1mm]
    &\times\bra{m\pr n\pr}\rho\ket{mn}\sum\limits_{i}\omega^{i(m-m\pr)}\sum\limits_{j}\omega^{j(n-n\pr)}\nonumber\\[1mm]
    & =\tfrac{d^{2}}{(\sum_{k}\lambda_{k})^{2}}\sum\limits_{m,n}\lambda_m\lambda_n\bra{mn}\rho\ket{mn}=:c_{\lambda},
    \label{eq:c lambda}
\end{align}
{where we have defined the normalization factor $c_{\lambda}$ as the inverse of the overall photon number and added the subscript $\lambda$ to emphasize the dependence on the initial measurements in the standard basis. If we had naively considered} the coincidence counts $\tilde{N}_{ij}$ in the tilted basis, and the quantity analogous to the right-hand side of Eq.~(\ref{eq:matrix elements in terms of coincidence counts appendix}), we would {have found} $\sum_{i,j}\tfrac{\tilde{N}_{ij}}{\sum_{k,l}\tilde{N}_{k,l}}=1$, by construction. To relate the coincidences to the matrix elements w.r.t. to the tilted basis, we {hence include} the additional normalization factor $c_{\lambda}$ of Eq.~(\ref{eq:c lambda}), i.e.,
\begin{align}
    \bra{\tilde{i}\tilde{j^{*}}}\rho\ket{\tilde{i}\tilde{j^{*}}}   &=\,c_{\lambda}\tfrac{\tilde{N}_{ij}}{\sum_{k,l}\tilde{N}_{k,l}},
\end{align}
as stated in the main text.

\subsection{Noise robustness}\label{sec:appendix noise robustness}

\begin{figure}[t!]
	\includegraphics[width=0.45\textwidth,trim={0cm 0mm 0cm 0mm}]{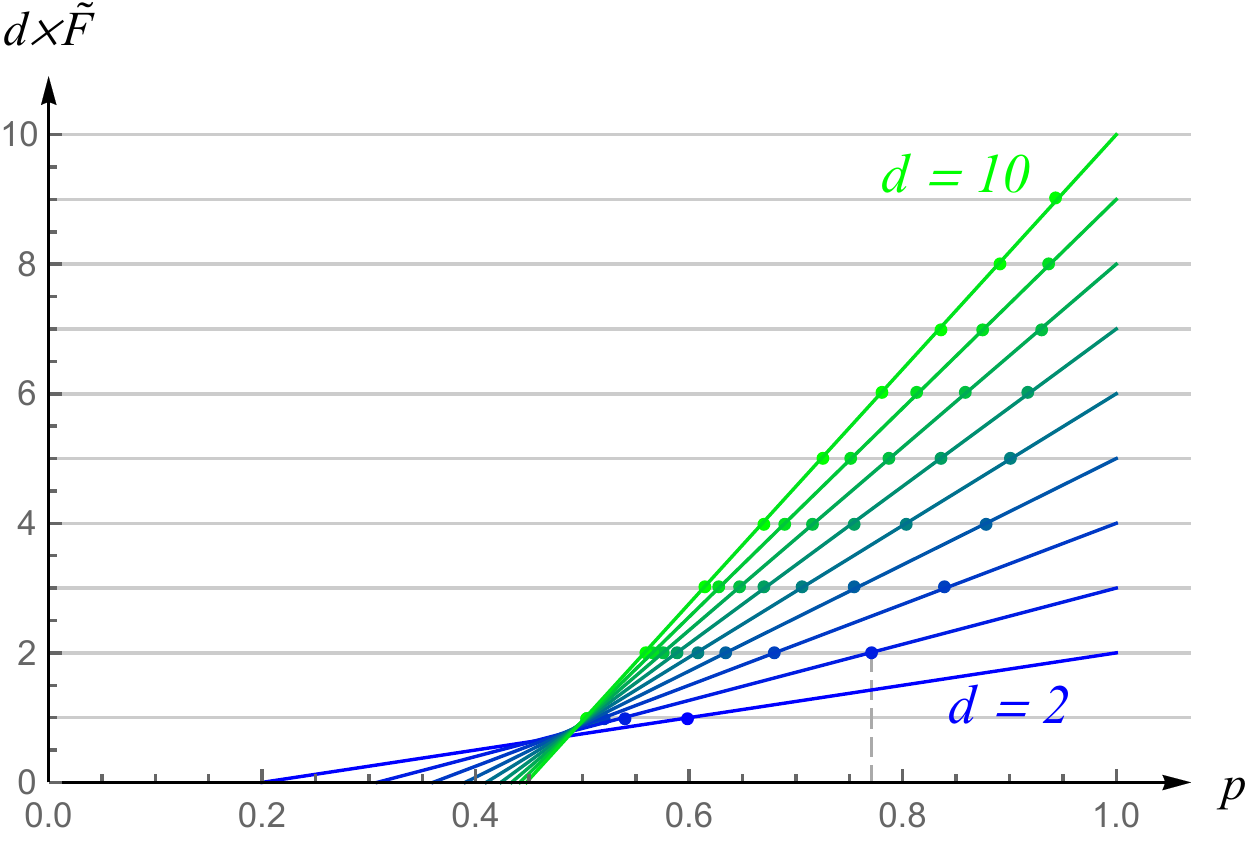}
\caption{\textbf{Noise-resistance of the fidelity bound for high-dimensional isotropic states.} The curves show {the fidelity bound $\tilde{F}(\rho_{\mathrm{iso}}(p),\Phi^{+})$ (weighted by the dimension $d$)} for isotropic states {$\rho_{\hspace*{0.5pt}\mathrm{iso}}(p)= p\ket{\hspace*{-0.5pt}\Phi^{+}\!}\!\!\bra{\!\Phi^{+}\!}+\tfrac{1-p}{d^{2}}\mathds{1}$} in $d\times d$ dimensions as functions of the visibility $p$ for $d=2$ (blue) to $d=10$ (green) in steps of $1$. The intersections of the curves with the horizontal lines at the points $\bigl(p_{k}(d),d\times\tilde{F}(\rho_{\mathrm{iso}}(p_{k}),\Phi^{+})\bigr)$ (colored dots), where the intersection coordinates on the vertical axis are
$d\times \tilde{F}(\rho_{\mathrm{iso}}(p_{k}),\Phi^{+})=d\times B_{k}(\Phi^{+})=k\in\{1,\ldots,9\}$, indicate that visibilities $p>p_{k}$ certify an entanglement dimensionality of at least $d_{\mathrm{ent}}=k+1$. In other words, for any $p$ the certified dimension is $d_{\mathrm{ent}}=\lceil d\times\tilde{F}(\rho_{\mathrm{iso}},\Phi^{+})\rceil$. For instance, for isotropic states in local dimension $d=3$ with visibility $p>p_{k=2}(d=3)=\tfrac{10}{13}$ (vertical dashed line), our fidelity bound certifies Schmidt rank $d_{\mathrm{ent}}=3$.}
\label{fig:dimensionality witnesses isotropic state}
\end{figure}

In this section, we discuss the special case of a maximally entangled target state, which is particularly interesting for several reasons. First, it provides a simple theoretical testing ground to evaluate the performance of our method in the presence of noise, as illustrated in Fig.~\ref{fig:dimensionality witnesses isotropic state}. There, we assume $\rho$ to be a mixture of $\ket{\Phi^{+}}$ with a maximally mixed state, i.e., an isotropic state $\rho_{\hspace*{0.5pt}\mathrm{iso}}= p\ket{\hspace*{-0.5pt}\Phi^{+}\!}\!\!\bra{\!\Phi^{+}\!}+\tfrac{1-p}{d^{2}}\mathds{1}$, where the visibility $p$ satisfies $0\leq p\leq1$ and $\mathds{1}$ is the identity operator in dimension $d^{2}$. This allows us to identify the visibility thresholds for the certification of the Schmidt ranks of maximally entangled states subject to white noise. Second, the fidelity bounds for the target state $\ket{\Phi^{+}}$ can be used to construct bounds on the entanglement of formation, as explained in the Supplementary Information. Although the selection of $\ket{\Phi^{+}}$ as a target state may not be optimally suited for a given experimental situation, it thus nonetheless provides an efficient method for the direct certification of the number of e-bits in the system. In Appendix~\ref{sec:appendix robustness of EoF bounds}, we show that this entanglement quantification method outperforms previous approaches~\cite{ErkerKrennHuber2017} in terms of detected e-bits and noise robustness.

\subsection{Improved bounds using multiple bases}\label{sec:appendix more than one tilted basis}

Next, we will show how measurements in more than one tilted basis can be included to improve the fidelity bounds. To this end, first note that the choice of tilted basis is not unique. For instance, all of the statements made so far about the properties of the tilted basis would remain unaffected if additional phase factors independent of $j$ were to be included in the definition of $\ket{\tilde{j}}$. That is, we have only relied on using identities such as $\sum_{j}\omega^{j(m-n)}=d\delta_{mn}$. For example, let us consider a family of tilted bases $\{\ket{\tilde{j}_{k}}\}_{j,k}$ parameterized by an integer $k\geq0$, such that
\begin{align}
    \ket{\tilde{j}_{k}} &=\,\frac{1}{\sqrt{\sum_{n}\lambda_{n}}} \sum_{m=0}^{d-1} \omega^{jm+km^{2}}\sqrt{\lambda_{m}} \ket{m}.
    \label{eq:tilted basis k appendix}
\end{align}
For $k=0$ we hence recover the original tilted basis. {When the target state is a product state (and hence separable), all vectors within any tilted basis collapse to the same standard basis vector (up to a global phase factor), and are hence fully contained within the standard basis. In this case, and indeed, whenever any of the Schmidt coefficients vanish identically, tilted bases are no longer complete, and hence cannot technically even be considered to be bases anymore. However, w}hen the target state is maximally entangled, $\ket{\Phi}=\ket{\Phi^{+}}$, we have $\lambda_{n}=\tfrac{1}{\sqrt{d}}\,\forall\,n$, in which case all of the tilted bases become orthonormal. Moreover, in this case one can recognize this construction as that of Ref.~\cite{WoottersFields1989}, i.e., for prime dimensions the choices $k=0,1,\ldots,d-1$ provide a maximal set of $d$ mutually unbiased bases (MUBs), $d+1$ if one includes the standard basis $\{\ket{m}\}_{m}$. For non-prime dimensions, the construction still provides an MUB w.r.t. to the standard basis for every choice of $k$, but the bases for different $k$ are in general not unbiased w.r.t to each other. We will return to these interesting special cases in Sec.~\ref{sec:appendix robustness of EoF bounds}.

{In the more realistic scenario where $\ket{\Phi}$ is not separable but also not maximally entangled and all Schmidt coefficients $\lambda_{m}$ (as estimated from initial measurements in the standard basis) have arbitrary nonzero values, we may construct nonorthogonal but complete tilted bases $\{\ket{\tilde{j}_{k}}\}_{j,k}$ according to Eq.}~(\ref{eq:tilted basis k appendix}){. As for the MUBs, this construction provides $d$ inequivalent tilted bases for odd prime dimensions, measurements w.r.t. which are sufficient for the fidelity bound to become tight, as we shall discuss in the following. To see this, first} note that the only contribution of the additional phases $\omega^{km^{2}}$ appears in the complex coefficient $c_{mnm\pr n\pr}=\sum_{j}\omega^{j(m-m\pr-n+n\pr)}$, which we can then replace by
\begin{align}
    c_{mnm\pr n\pr}\suptiny{0}{0}{(k)}:=\sum_{j}\omega^{j(m-m\pr-n+n\pr)}\,\omega^{k(m^{2}-m^{\prime\hspace*{0.5pt}2}-n^{2}+n^{\prime\hspace*{0.5pt}2})}.
\end{align}
Clearly, when using any single one of the bases $\{\ket{\tilde{j}_{k}}\}_{j,k}$, the modification of the constant $c_{mnm\pr n\pr}$ becomes irrelevant again due to the modulus, i.e., $|c_{mnm\pr n\pr}\suptiny{0}{0}{(k)}|=|c_{mnm\pr n\pr}\suptiny{0}{0}{(0)}|$ for all $k$.

However, we may use several of the tilted bases simultaneously to obtain an advantage. Replacing the term $\Sigma=\sum\limits_{j=0}^{d-1}\bra{\tilde{j}\tilde{j}^{*}}\rho\ket{\tilde{j}\tilde{j}^{*}}$ by an average over $M$ different tilted bases as defined by Eq.~(\ref{eq:tilted basis k appendix}), i.e.,
\begin{align}
    \Sigma\rightarrow\Sigma\suptiny{0}{0}{(M)}  &=\,
    \tfrac{1}{M}\sum\limits_{k=0}^{M-1}
    \sum\limits_{j=0}^{d-1}\bra{\tilde{j}_{k}\tilde{j}_{k}^{*}}\rho\ket{\tilde{j}_{k}\tilde{j}_{k}^{*}},
\end{align}
one finds that the only affected term in the bound $\tilde{F}_{2}$ for $F_{2}$ is $\Sigma_{3}$. That is, we may replace the coefficient $\tilde{\gamma}_{mm\pr nn\pr}$ by the modified coefficient
\begin{align}
    \tilde{\gamma}_{mm\pr nn\pr}\suptiny{0}{0}{(M)}  &=
    \tilde{\gamma}_{mm\pr nn\pr}\,\tfrac{1}{M}\left|\sum\limits_{k=0}^{M-1}\omega^{k(m^{2}-m^{\prime\hspace*{0.5pt}2}-n^{2}+n^{\prime\hspace*{0.5pt}2})}\right|,
    \label{eq:tilde gamma M appendix}
\end{align}
and define the quantity $\tilde{F}\suptiny{0}{0}{(M)}:=F_{1}+\tilde{F}_{2}\suptiny{0}{0}{(M)}\leq F$, where
\begin{align}
    \tilde{F}_{2}\suptiny{0}{0}{(M)}    &:=\tfrac{\left(\sum_m\lambda_m\right)^2}{d}\Sigma\suptiny{0}{0}{(M)}
        -\!\!\!\sum_{m,n=0}^{d-1}\!\! \lambda_m\lambda_n \bra{mn}\rho\ket{mn}  \nonumber\\
	&-\hspace*{-4.5mm}
    \sum\limits_{\substack{m\neq m\pr\!,  m\neq n \\ n\neq n\pr\!, n\pr\neq m\pr}}\hspace*{-4.5mm}
    \tilde{\gamma}_{mm\pr nn\pr}\suptiny{0}{0}{(M)} \,
    \sqrt{\bra{m\pr n\pr}\rho\ket{m\pr n\pr}\bra{mn}\rho\ket{mn}}.
    \label{eq:F2 bound tilted basis appendix improved}
\end{align}
In the least favourable possible case all phases in the sum over $k$ are aligned and $\tilde{\gamma}_{mm\pr nn\pr}\suptiny{0}{0}{(M)}=\tilde{\gamma}_{mm\pr nn\pr}$, but in general $\tilde{\gamma}_{mm\pr nn\pr}\suptiny{0}{0}{(M)}\leq\tilde{\gamma}_{mm\pr nn\pr}$. Consequently, the fidelity bounds can only be improved by including measurements in more than one tilted basis.

In fact, when the dimension $d$ is a (non-even) prime, we have $\tilde{F}\suptiny{0}{0}{(M\pr)}\geq\tilde{F}\suptiny{0}{0}{(M)}$ for $M\pr\geq M$, and for $M=d$ the prefactor $\tilde{\gamma}_{mm\pr nn\pr}\suptiny{0}{0}{(M=d)}$ vanishes exactly and the fidelity bound becomes tight, i.e., $F=\tilde{F}\suptiny{0}{0}{(M=d)}$ . In order to show this, we need to examine the sum in Eq.~(\ref{eq:tilde gamma M appendix}). At first it is important to realize that since the value of $\tilde{\gamma}_{mm\pr nn\pr}$ does not depend on $k$, only cases for which $(m-m\pr-n+n\pr) \mod d = 0$ need to be examined, otherwise $\tilde{\gamma}_{mm\pr nn\pr} = 0$ leads to $\tilde{\gamma}_{mm\pr nn\pr}\suptiny{0}{0}{(M)} = 0$. Let us therefore prove the following claim. For parameter choices fulfilling the conditions
\begin{align}
    &m\neq m\pr, m\neq n, \nonumber\\
    &n\neq n\pr, n\pr\neq m\pr, \nonumber\\
    &(m - m\pr -n + n\pr) \mod d = 0
    \label{eq:condition3}
\end{align}
it holds that $(m^{2}-m^{\prime\hspace*{0.5pt}2}-n^{2}+n^{\prime\hspace*{0.5pt}2}) \neq 0$. We will prove this claim by contradiction. In order to do so, suppose that both of the following equalities hold
\begin{align}
    m+n\pr &= m\pr + n \mod d\\
    m^2 + (n\pr)^2 &= (m\pr)^2 + n^2 \mod d.
    \label{eq:squares}
\end{align}
Without loss of generality suppose $m>n$, which also implies $m\pr > n\pr$. Let us define $c:= m-n = m\pr-n\pr$, which allows us to rewrite Eq.~(\ref{eq:squares}) as
\begin{align}
    m^2 + n^{\prime\hspace*{0.5pt}2} &= (n\pr +c)^2 + (m-c)^2 \mod d\nonumber\\
    m^2 + n^{\prime\hspace*{0.5pt}2}&= (n\pr)^2 + 2cn\pr + c^2  + m^2 - 2cm +c^2 \mod d\nonumber\\
    0&=  2c^2 + 2cn\pr - 2cm  \mod d\nonumber\\
    0&=  2c (c+ n\pr - m)  \mod d\nonumber\\
    0&=  2c (m\pr - m)  \mod d
    \label{eq:contradiction}.
\end{align}
The last equality holds, if and only if $2c(m\pr-m)$ is a multiple of $d$. Since $d$ is an odd prime, the only possibility is that either $c$ or $(m\pr-m)$ are multiples of $d$. Clearly, since $c = m-n$, $m>n$ and $m,n \in \{0,\dots, d-1\}$, $0<c<d$, and $c$ is therefore not a multiple of $d$. Similarly, since $m\neq m\pr$ and $m,m\pr \in \{0,\dots,d-1\}$, $-d<(m\pr-m)<d$, therefore $(m\pr-m)$ is not a multiple of $d$. We hence arrive at a contradiction with Eq.~(\ref{eq:contradiction}) and conclude that under the conditions of~(\ref{eq:condition3}) we have $(m^{2}-m^{\prime\hspace*{0.5pt}2}-n^{2}+n^{\prime\hspace*{0.5pt}2}) \neq 0$.

Therefore, when working with $M$ different tilted bases, $\sum_{k=0}^{M-1} \omega^{k(m^{2}-m^{\prime\hspace*{0.5pt}2}-n^{2}+n^{\prime\hspace*{0.5pt}2})}$ is a sum of $M$ \emph{different}\footnote{The difference of the powers results from the fact that in the mod prime multiplicative group, every non-zero element is a generator of the whole group. This means that since $(m^2 - m^{\prime\hspace*{0.5pt}2} - n^2 + n^{\prime\hspace*{0.5pt}2})$ is non-zero, iterating over different values of $k$ results in different values of the whole exponent.} powers of $\omega$. We subsequently have to show that the absolute value of this sum can be bounded to be strictly lower than $M$. Moreover, the bound improves with increasing $M$, and whenever $M=d$, the sum in Eq.~(\ref{eq:tilde gamma M appendix}) [and hence also the sum in the last line of Eq.~(\ref{eq:F2 bound tilted basis appendix improved})] vanishes. Before we turn to the more general statement for arbitrary $M$, let us briefly focus on the case $M=d$, where it can be easily seen that for non-zero $(m^2 - m^{\prime\hspace*{0.5pt}2} - n^2 + n^{\prime\hspace*{0.5pt}2})$
$\sum_{k=0}^{d-1} \omega^{k(m^2 - m^{\prime\hspace*{0.5pt}2} - n^2 + n^{\prime\hspace*{0.5pt}2}) }=0$.

For general values $M<d$ let us now analytically bound $\vert\sum_{k=0}^{M-1} \omega^{kc}\vert$, where $c$ is a non-zero {integer}. Naturally, the exact value of this sum depends on the particular value of $c$, but here we give a general bound. To this end, we first argue that the worst case (the highest possible sum) corresponds to the situation, where $kc$ ranges over subsequent powers of $\omega$ (i.e. $c=1$). This can be seen from the fact that powers of $\omega$ can be represented in the complex plane as vectors lying on the unit circle with the centre at the origin. The absolute value of the sum of several different powers of $\omega$ can therefore be seen as the size of the sum of their corresponding vectors. Recall that for odd-prime dimension $d$, the exponent $kc$ ranges over $M$ different numbers between $0$ and $d-1$. Now it is not hard to see that by fixing the number of vectors $M$, the worst case sum (i.e., the largest absolute value) corresponds to the sum of the $M$ vectors next to each other on the complex plane, which in turn corresponds to the subsequent powers of $\omega$. With this knowledge, we have to bound one particular worst case sum, given by
\begin{align}
    \sum_{k=0}^{M-1} \omega^k = \sum_{k=0}^{M-1}e^{\frac{2\pi i k}{d}}.
\end{align}
Using a variant of the Dirichlet kernel~\cite{Dirichlet1829}, i.e.,
\begin{align}
    \sum_{k=0}^{M-1} e^{iMx} = e^{\frac{i(M-1)x}{2}}\frac{\sin{\left(\frac{Mx}{2}\right)}}{\sin{\left(\frac{x}{2}\right)}}
\end{align}
with $x = \frac{2\pi}{d}$, we have
\begin{align}
    \sum_{k=0}^{M-1} \omega^k = e^{\frac{i(M-1)\pi}{d}}\frac{\sin{\left(\frac{M\pi}{d}\right)}}{\sin{\left(\frac{\pi}{d}\right)}}.
\end{align}
\begin{figure}[ht!]
	\includegraphics[width=0.49\textwidth,trim={0cm 0mm 0cm 0mm}]{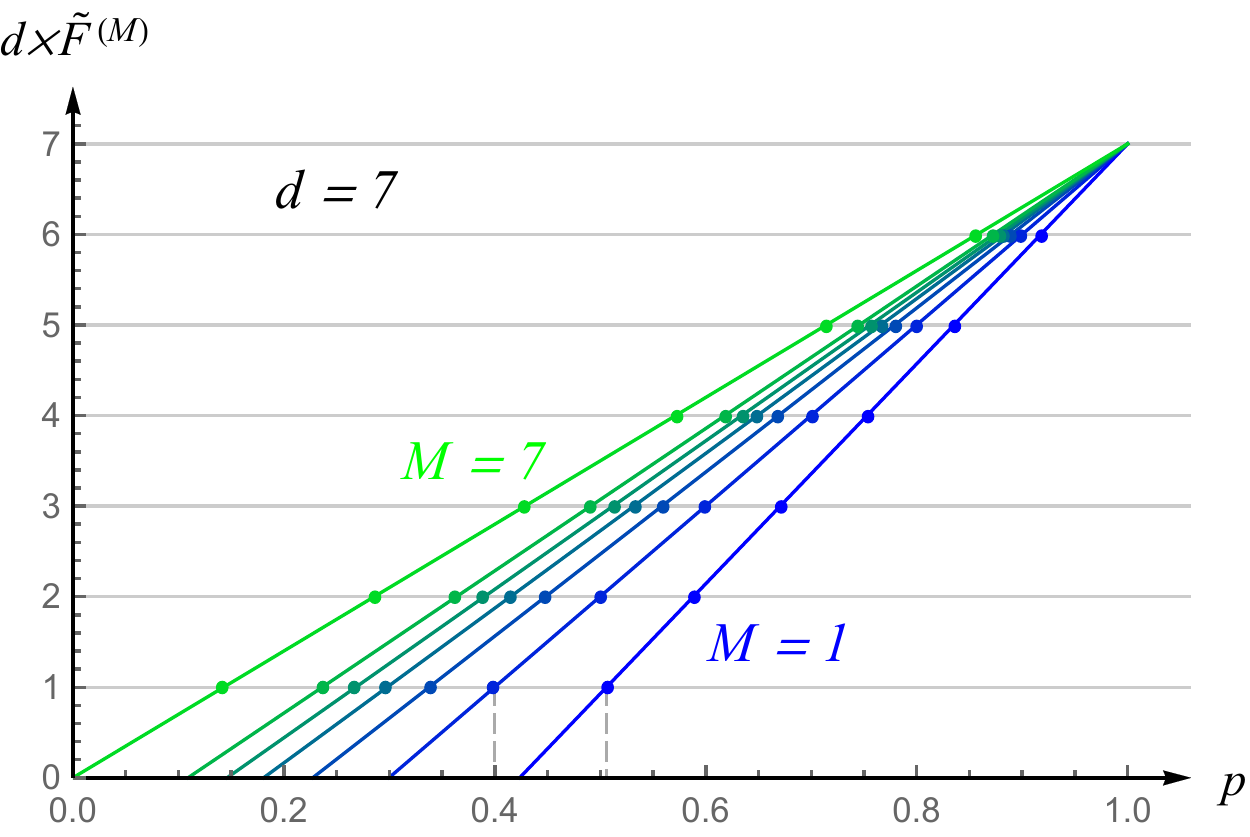}
\vspace*{-2mm}
\caption{\textbf{Improved {fidelity bound \&} dimensionality witness for isotropic state.} The curves show the fidelity bound $\tilde{F}\suptiny{0}{0}{(M)}(\rho_{\mathrm{iso}}(p),\Phi^{+})$ (weighted by the local dimension $d=7$) for isotropic states $\rho_{\hspace*{0.5pt}\mathrm{iso}}(p)= p\ket{\hspace*{-0.5pt}\Phi^{+}\!}\!\!\bra{\!\Phi^{+}\!}+\tfrac{1-p}{d^{2}}\mathds{1}$ in $d\times d$ dimensions as functions of the visibility $p$ for local dimension $d=7$ for different numbers of global product bases, i.e., $M=1$ (blue) to $M=7$ (green) in steps of~$1$. The intersections of the curves with the horizontal lines at the points $\bigl(p_{k}\suptiny{0}{0}{(M)}(d),d\times\tilde{F}\suptiny{0}{0}{(M)}(\rho_{\mathrm{iso}}(p_{k}\suptiny{0}{0}{(M)}),\Phi^{+})\bigr)$ (colored dots), where the intersection coordinates on the vertical axis are $d\times \tilde{F}\suptiny{0}{0}{(M)}(\rho_{\mathrm{iso}}(p_{k}\suptiny{0}{0}{(M)}),\Phi^{+})=d\times B_{k}(\Phi^{+})=k\in\{1,\ldots,6\}$, indicate that visibilities $p>p_{k}\suptiny{0}{0}{(M)}$ certify an entanglement dimensionality of at least $d_{\mathrm{ent}}=k+1$ when $M$ tilted bases are used. In other words, for any $p$ the certified dimension is $d_{\mathrm{ent}}=\lceil d\times\tilde{F}\suptiny{0}{0}{(M)}(\rho_{\mathrm{iso}},\Phi^{+})\rceil$. For instance, for isotropic states in local dimension $d=7$ our fidelity bound with one tilted basis ($M=1$) certifies Schmidt rank $d_{\mathrm{ent}}=2$ for a visibility $p>p_{k=1}\suptiny{0}{0}{(M=1)}(d=7)=\tfrac{43}{85}$ (right vertical dashed line), whereas for two tilted bases ($M=2$) a visibility $p>p_{k=1}\suptiny{0}{0}{(M=2)}(d=7)\approx0.3997$ (left vertical dashed line) is enough to certify $d_{\mathrm{ent}}=2$.}
\label{fig:dimensionality witnesses multiple bases isotropic state}
\end{figure}

\noindent
Taking the absolute value reveals that for any choice of non-zero {integer} $c$ we have
\begin{align}
    \left\vert\sum_{k=0}^{M-1} \omega^{kc}\right\vert \leq \frac{\left\vert{\sin\left(\frac{M\pi}{d}\right)}\right\vert}{\left\vert{\sin\left(\frac{\pi}{d}\right)}\right\vert}.
\end{align}
After plugging this lower bound into Eq. (\ref{eq:tilde gamma M appendix}), all (non-zero) prefactors $\tilde{\gamma}_{mm\pr nn\pr}\suptiny{0}{0}{(M)}$ become
decreasing functions of $M$, on the interval $1\leq M \leq d$, which concludes the proof that
$\tilde{F}\suptiny{0}{0}{(M\pr)}\geq\tilde{F}\suptiny{0}{0}{(M)}$ for $M\pr\geq M$ in odd prime dimensions.

For general dimension $d$, however, it is not the case that $\tilde{F}\suptiny{0}{0}{(M\pr)}\geq\tilde{F}\suptiny{0}{0}{(M)}$ for $M\pr\geq M$, except for the case
when $M=1$ (for any dimension). \\

An illustration of the improvement obtained by including multiple tilted bases is given in Fig.~\ref{fig:dimensionality witnesses multiple bases isotropic state} for an isotropic state $\rho_{\hspace*{0.5pt}\mathrm{iso}}= p\ket{\hspace*{-0.5pt}\Phi^{+}\!}\!\!\bra{\!\Phi^{+}\!}+\tfrac{1-p}{d^{2}}\mathds{1}$ in dimension $d=7$. Such a state{ }highlights the influence of white noise on the certification method, since the isotropic state is a mixture of a maximally entangled and a maximally mixed state. We have hence shown that an improvement of the bounds by using more than two {global product} bases is possible in principle. In Sec.~\ref{sec:appendix robustness of EoF bounds} we will further illustrate this improvement for quantifying entanglement.




\subsection{Bounds on the entanglement of formation}\label{sec:appendix EoF bounds}

In this section, we discuss a method for bounding the entanglement of formation in bipartite systems of arbitrary dimension. To provide a self-contained approach, let us first give a pedagogical review of the entanglement of formation and useful bounds for it also discussed in Ref.~\cite{ErkerKrennHuber2017}, before we make use of the fidelity bounds established thus far in Sec.~\ref{sec:appendix robustness of EoF bounds}. To begin, recall that the subsystems $A$ and $B$ of a pure bipartite state $\ket{\psi}\subtiny{-1}{0}{AB}$ are entangled if and only if their reduced states $\rho\subtiny{-1}{0}{A}=\tr\subtiny{-1}{0}{B}(\ket{\psi}\!\!\bra{\psi})$ and $\rho\subtiny{-1}{0}{B}=\tr\subtiny{-1}{0}{A}(\ket{\psi}\!\!\bra{\psi})$ are mixed. This fact can easily be seen from the Schmidt decomposition, i.e., that any pure state $\ket{\psi}\subtiny{-1}{0}{AB} \in \mathcal{H}\subtiny{-1}{0}{AB} = \mathcal{H}\subtiny{-1}{0}{A} \otimes \mathcal{H}\subtiny{-1}{0}{B}$ may be written as
\begin{align}
    \ket{\psi}\subtiny{-1}{0}{AB}   &=\,\sum\limits_{m=0}^{k-1}\lambda_{m}\ket{\phi_{m}}\subtiny{-1}{0}{A}\ket{\chi_{m}}\subtiny{-1}{0}{B}
\end{align}
with respect to the Schmidt bases $\{\ket{\phi_{m}}\subtiny{-1}{0}{A}\}_{m}$ and $\{\ket{\chi_{m}}\subtiny{-1}{0}{B}\}_{m}$, and where $k\leq\min\{\dim(\mathcal{H}\subtiny{-1}{0}{A}),\dim(\mathcal{H}\subtiny{-1}{0}{B})\}$. The entanglement of the state $\ket{\psi}\subtiny{-1}{0}{AB}$ may therefore be quantified by the mixedness $1-\tr(\rho\subtiny{-1}{0}{A}^{2})$ of the reduced states. More specifically, we can define the \emph{entropy of entanglement} $\mathcal{E}_{\mathrm{L}}$ via the linear entropy $S_{\mathrm{L}}$ as
\begin{align}
    \mathcal{E}_{\mathrm{L}}(\ket{\psi}) &=\,S_{\mathrm{L}}(\rho\subtiny{-1}{0}{A})  \,=\,\sqrt{2\bigl(1-\tr(\rho\subtiny{-1}{0}{A}^{2})\bigr)}.
\end{align}
This method for entanglement quantification can be extended to mixed states via a convex-roof construction, i.e., we define
\begin{align}
    \mathcal{E}_{\mathrm{L}}(\rho) &:=\,\inf_{\mathcal{D}(\rho)}\sum\limits_{i}p_{i}\,S_{\mathrm{L}}(\rho\subtiny{-1}{0}{A}\suptiny{0}{0}{(i)})\,,
\end{align}
where the infimum is taken over the set of all pure state decompositions of $\rho$, i.e.,
\begin{align}
\mathcal{D}(\rho)   &=\,\Bigl\{\!\left\{\bigl(p_{i},\psi_{i}\bigr)\right\}_{i}|\rho\!=\!\sum\limits_{i}p_{i}\ket{\psi_{i}}\!\!\bra{\psi_{i}}, 0\leq p_{i}\leq\!1, \!\sum\limits_{i}p_{i}\!=\!1\!\Bigr\},
\end{align}
where $\rho\subtiny{-1}{0}{A}\suptiny{0}{0}{(i)}=\tr\subtiny{-1}{0}{B}\bigl(\ket{\psi_{i}}\!\!\bra{\psi_{i}}\bigr)$.\\

A simple bound on this convex roof of the linear entropy was derived in Refs.~\cite{HuberDeVicente2013, HuberPerarnauLlobetDeVicente2013}. Defining the quantity
\begin{align}
    I(\rho) &=\,\sqrt{\tfrac{2}{d(d-1)}}\sum\limits_{m\neq n}\Bigl(
    \left|\bra{mm}\rho\ket{nn}\right|\nonumber\\[1mm]
    &\ - \sqrt{\bra{mn}\rho\ket{mn}\!\bra{nm}\rho\ket{nm}}\Bigr),
\end{align}
for bipartite systems of equal local dimension $d$, i.e., $\dim(\mathcal{H}\subtiny{-1}{0}{A})=\dim(\mathcal{H}\subtiny{-1}{0}{B})=d$, with bases $\{\ket{\phi_{n}}\subtiny{-1}{0}{A}\equiv\ket{n}\subtiny{-1}{0}{A}\}$ and $\{\ket{\chi_{n}}\subtiny{-1}{0}{B}\equiv\ket{n}\subtiny{-1}{0}{B}\}$, it was shown in~\cite{HuberDeVicente2013, HuberPerarnauLlobetDeVicente2013} that
\begin{align}
    I(\rho) &   \leq\,\mathcal{E}_{\mathrm{L}}(\rho)\,.
\end{align}
Now, we want to see how $I(\rho)$ can used to bound also the entanglement of formation (EoF)~\cite{BennettDiVincenzoSmolinWootters1996,Wootters1998}, defined as the convex roof extension of the entropy of entanglement when the von Neumann entropy $S(\rho)=-\tr\bigl(\rho\log(\rho)\bigr)$ is used instead of the linear entropy, i.e.,
\begin{align}
    \mathcal{E}_{\mathrm{oF}}(\rho) &:=\,\inf_{\mathcal{D}(\rho)}\sum\limits_{i}p_{i}\,S(\rho\subtiny{-1}{0}{A}\suptiny{0}{0}{(i)})\,.
\end{align}
To understand this connection, let us briefly expand upon the derivation given in Ref.~\cite{ErkerKrennHuber2017}. First, note that for pure states $\ket{\psi}$ we have
\begin{align}
    I(\ket{\psi}) &\leq\,\mathcal{E}_{\mathrm{L}}(\ket{\psi})   \,=\,\sqrt{2\bigl(1-\tr(\rho\subtiny{-1}{0}{A}^{2})\bigr)}\,.
\end{align}
Therefore, if $I(\ket{\psi})\geq0$ we can write
\begin{align}
    \tr(\rho\subtiny{-1}{0}{A}^{2}) &\leq\,1\,-\,\tfrac{1}{2}I^{2}(\ket{\psi}),
\end{align}
which implies that
\begin{align}
    -\log\bigl(\tr(\rho\subtiny{-1}{0}{A}^{2})\bigr) &\geq\,-\,\log\Bigl(1\,-\,\tfrac{1}{2}I^{2}(\ket{\psi})\Bigr)
\end{align}
since $\log{x}$ is a monotonically increasing function. With the additional negative sign we can recognize the left-hand side as the R{\'e}nyi $2$-entropy, defined as
\begin{align}
    S_{\alpha}(\rho)    &:=\,\frac{1}{1-\alpha}\log\tr(\rho^{\alpha})
    \label{eq:Renyi entropy}
\end{align}
for $\alpha=2$. For all $\alpha,\beta\in\mathbb{N}$ and for all $\rho$, the R{\'e}nyi entropies satisfy $S_{\alpha}(\rho)\geq S_{\beta}(\rho)$ for $\alpha\leq\beta$. In particular,
this means that
\begin{align}
    S_{1}(\rho) &=\,\lim_{\alpha\rightarrow1}S_{\alpha}(\rho) \,\geq\,
    S_{2}(\rho) \,=\,-\log\bigl(\tr(\rho^{2})\bigr)
\end{align}
and consequently one has
\begin{align}
    S_{1}(\rho\subtiny{-1}{0}{A})   &\geq\,-\,\log\Bigl(1\,-\,\tfrac{1}{2}I^{2}(\ket{\psi})\Bigr).
\end{align}
For pure states, the (von Neumann) entropy of the subsystem is equal to the EoF and we have hence obtained the desired bound. To see that the bound also holds for mixed states, simply note that $-\log(1-x^{2}/2)$ is a convex function. Similarly, the function $I(\rho)$ is convex, since
\begin{align}
    I_{1}   &:=\sum\limits_{m\neq n}\left|\bra{mm}\rho\ket{nn}\right|
    \label{eq: I1}
\end{align}
is convex, while
\begin{align}
    I_{2}   &:=\sum\limits_{m\neq n}\sqrt{\bra{mn}\rho\ket{mn}\!\bra{nm}\rho\ket{nm}}
\end{align}
is concave, i.e., by Jensen's inequality~\cite{Jensen1906}
\begin{align}
    I_{1}(\sum\limits_{i}p_{i}\rho_{i}) &\leq\,\sum\limits_{i}p_{i}I_{1}(\rho_{i}),\\[1mm]
    I_{2}(\sum\limits_{i}p_{i}\rho_{i}) &\geq\,\sum\limits_{i}p_{i}I_{2}(\rho_{i}),
\end{align}
for $0\leq p_{i}\leq1$ and $\sum_{i}p_{i}=1$. This allows us to conclude that for all states $\rho$, for which $I(\rho)\geq0$ one has
\begin{align}
    \mathcal{E}_{\mathrm{oF}}(\rho)    &\geq\,-\,\log\Bigl(1\,-\,\tfrac{1}{2}I^{2}(\rho)\Bigr).
\end{align}

Here, it is first useful to note here that the value of $I(\rho)$ (in particular, whether or not $I$ is non-negative) for a given state depends on the bases $\{\ket{m}\subtiny{-1}{0}{A}\}_{m}$ and $\{\ket{n}\subtiny{-1}{0}{B}\}_{n}$ that are chosen. For instance, if both bases are chosen to be the same single-qubit bases 
 and the quantum state in question is the singlet state $\ket{\psi^{-}}=\tfrac{1}{\sqrt{2}}\bigl(\ket{01}-\ket{10}\bigr)$, where $\ket{0}$ and $\ket{1}$ are assumed to be the eigenstates of the third Pauli matrix $Z=\diag\{1,-1\}$, then $I(\ket{\psi^{-}})=-1$. In other words, the bases $\{\ket{m}\subtiny{-1}{0}{A}\}_{m}$ and $\{\ket{n}\subtiny{-1}{0}{B}\}_{n}$ should be chosen with a specific family of states in mind. For pure states, it is most useful to choose the Schmidt bases of the two subsystems.\\

Second, observe that, on the one hand, the term $I_{2}$ contains only diagonal matrix elements and {hence can practically easily be} estimated using measurements in one pair of {global product} bases only. That is, counting the coincidences $N_{mn}$ in the basis setting $\ket{m}\subtiny{-1}{0}{A}\ket{n}\subtiny{-1}{0}{B}$, we can reconstruct the desired matrix elements as $\bra{mn}\rho\ket{mn}=N_{mn}/\bigl(\sum_{i,j}N_{ij}\bigr)$. On the other hand, to estimate the off-diagonal matrix elements of the term $I_{1}$ precisely, one would be required to reconstruct the entire density matrix by way of state tomography. However, this costly procedure can be avoided by supplementing the measurements in the basis $\{\ket{mn}\}_{m,n}$ by measurements in one (or more) MUBs w.r.t. $\{\ket{mn}\}_{m,n}$ to provide a lower bound on $I_{2}(\rho)$.


\subsection{Entanglement quantification using mututally unbiased bases}\label{sec:appendix robustness of EoF bounds}

Having established the usefulness of the quantity $I(\rho)$ for bounding the entanglement of formation, let us now relate it to the fidelity bounds we have discussed before. Inspection of the fidelity to the maximally entangled state, i.e.,
\begin{align}
    F(\rho,\Phi^{+})    &=\,\tfrac{1}{d}\sum\limits_{m}\bra{mm}\rho\ket{mm}\,+\,\tfrac{1}{d}\sum\limits_{m\neq n}\bra{mm}\rho\ket{nn},
\end{align}

\begin{figure}[ht!]
	(a)\includegraphics[width=0.47\textwidth,trim={0cm 0mm 0cm 0mm}]{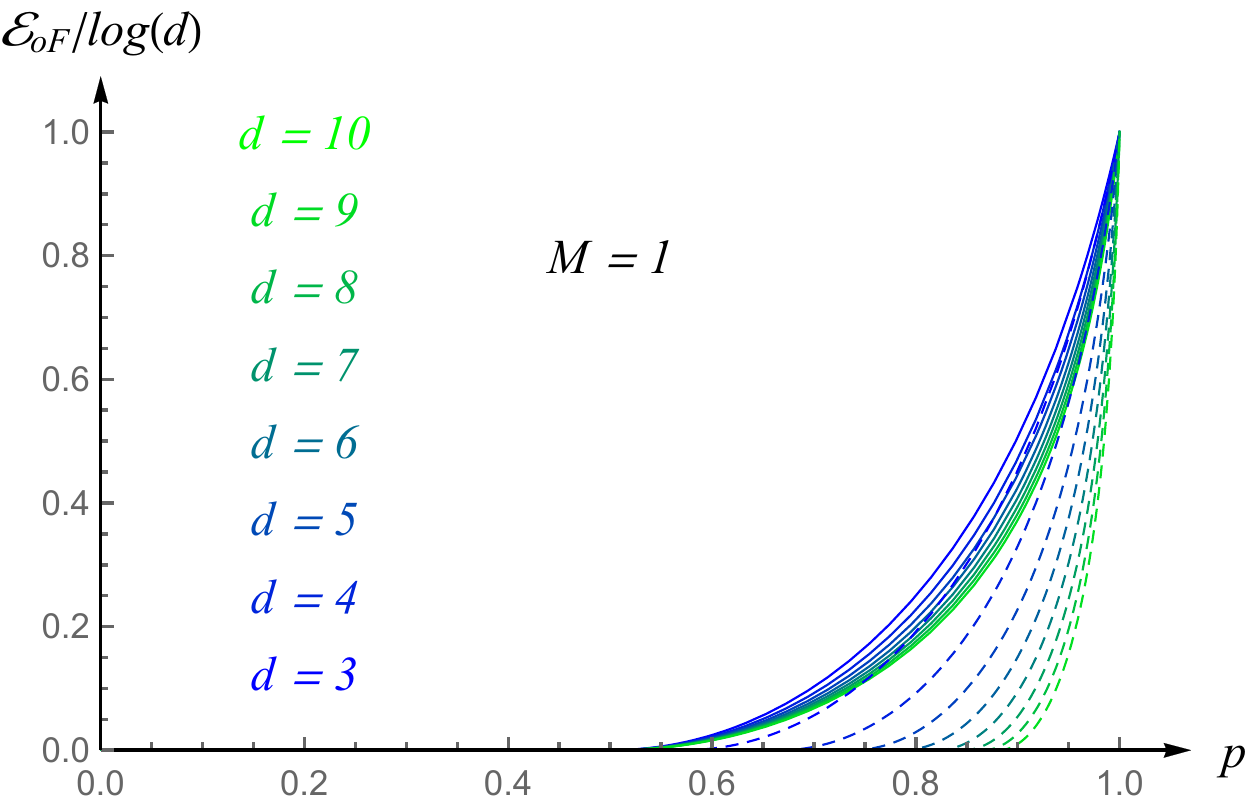}
	(b)\includegraphics[width=0.47\textwidth,trim={0cm 0mm 0cm 0mm}]{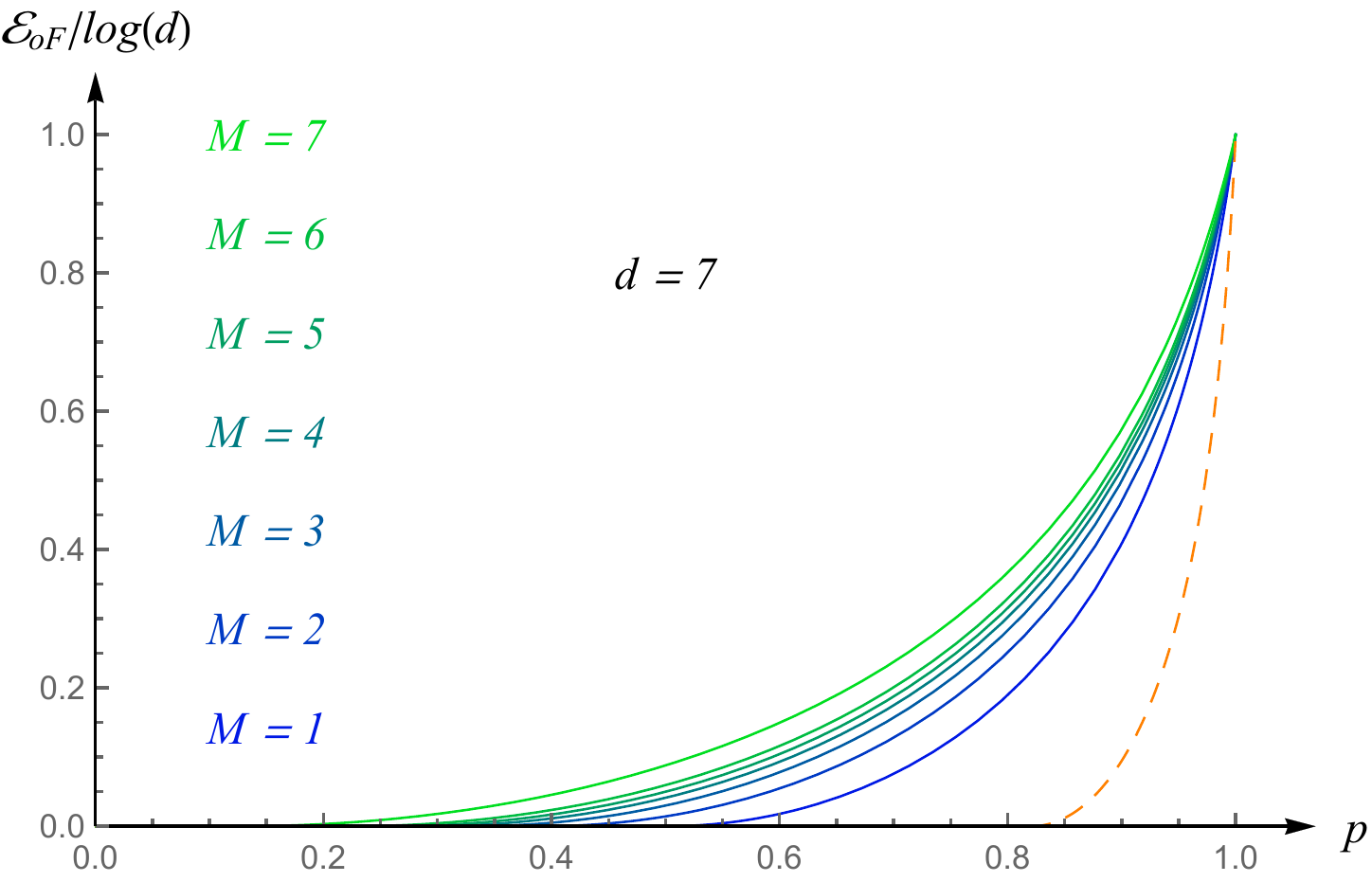}
\caption{\textbf{Entanglement bounds for isotropic state.} (a) The dashed and solid curves show the lower bounds for $\mathcal{E}_{\mathrm{oF}}$ obtained for $M=1$ and $\rho_{\hspace*{0.5pt}\mathrm{iso}}(p)$ using the bounds from Ref.~\cite{ErkerKrennHuber2017} (dashed) and using the bound presented here in (\ref{eq:quitter bound for I}) (solid curves), respectively, for dimensions $d=3$ (blue) to $d=10$ (green) in steps of $1$ and in units of $\log{d}$. It can be seen that the newly improved bounds can certify higher entanglement for given visibilities $p$. (b) The bound of Ref.~\cite{ErkerKrennHuber2017} (orange, dashed) is compared with the bound of~(\ref{eq:quitter bound for I}) (solid curves) for fixed dimension $d=7$ and varying numbers of bases, $M=1$ (blue) to $M=7$ (green) in steps of $1$.}
\label{fig:EoF isotropic state}
\end{figure}

\noindent
immediately lets us obtain the bound
\begin{align}
    \sum\limits_{m\neq n}\left|\bra{mm}\rho\ket{nn}\right|  &\geq\,\sum\limits_{m\neq n}\bra{mm}\rho\ket{nn}\\[1mm]
    &\ =\,d\,F(\rho,\Phi^{+})-\sum\limits_{m}\bra{mm}\rho\ket{mm}.\nonumber
\end{align}
Since $F(\rho,\Phi^{+})\geq\tilde{F}\suptiny{0}{0}{(M)}$, this, in turn, implies that
\begin{align}
    &I(\rho) \geq
    \sqrt{\tfrac{2}{d(d-1)}}\Bigl(d\,\tilde{F}\suptiny{0}{0}{(M)}(\rho,\Phi^{+})-\sum\limits_{m}\bra{mm}\rho\ket{mm}
    \nonumber\\[1mm]
    &\ \ \ \ \ - \sum\limits_{m\neq n}\sqrt{\bra{mn}\rho\ket{mn}\!\bra{nm}\rho\ket{nm}}\Bigr)\label{eq:quitter bound for I}\\[1mm]
    &\geq\,\sqrt{\tfrac{2}{d(d-1)}}\Bigl(d\,\Sigma\suptiny{0}{0}{(M)}-1-\!\!
    \sum\limits_{m\neq n}\!\sqrt{\bra{mn}\rho\ket{mn}\!\bra{nm}\rho\ket{nm}}
    \nonumber\\[1mm]
    &\ \ \ -
    \sum\limits_{\substack{m\neq m\pr\!,  m\neq n \\ n\neq n\pr\!, n\pr\neq m\pr}}\hspace*{-4.5mm}
    \tilde{\gamma}_{mm\pr nn\pr}\suptiny{0}{0}{(M)} \,
    \sqrt{\bra{m\pr n\pr}\rho\ket{m\pr n\pr}\bra{mn}\rho\ket{mn}}
    \Bigr),\nonumber
\end{align}

\begin{figure}[ht!]
	\includegraphics[width=0.47\textwidth,trim={0cm 0mm 0cm 0mm}]{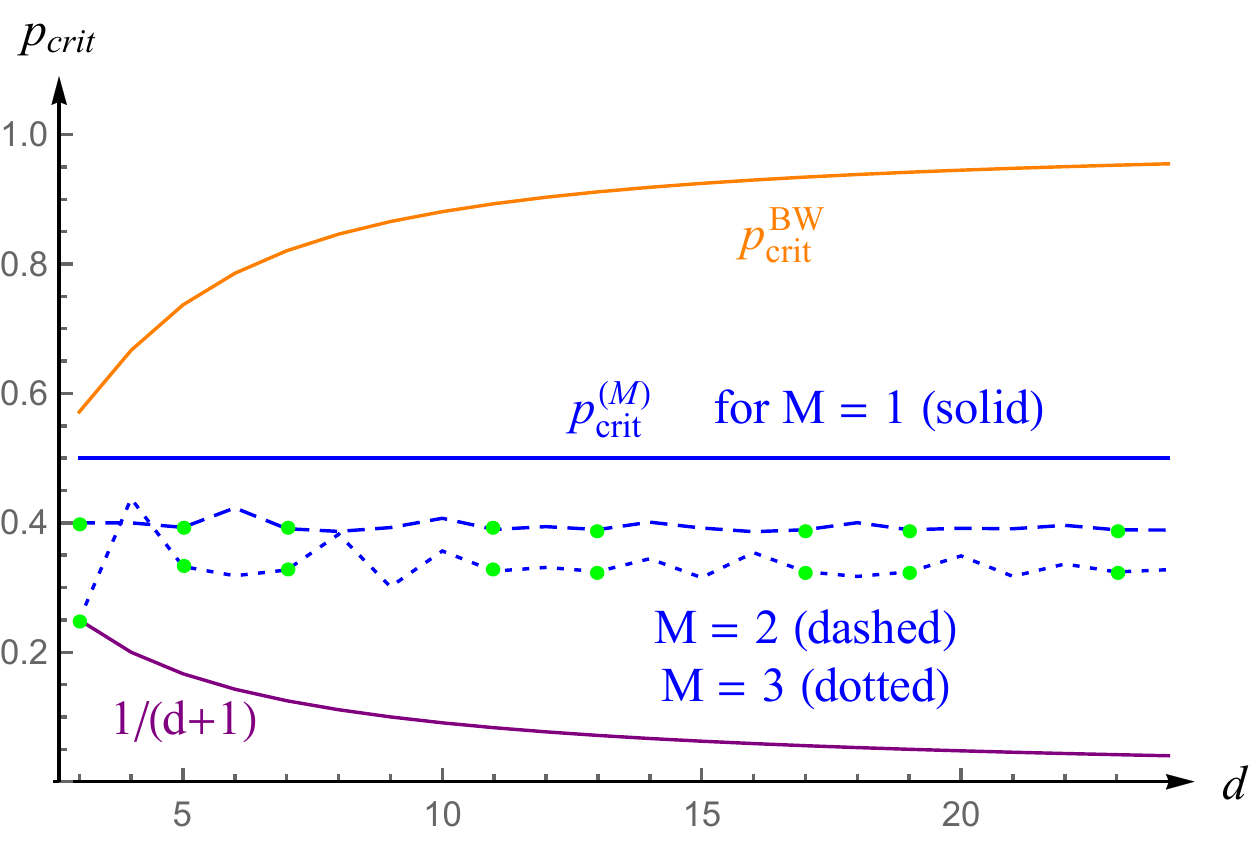}
\caption{\textbf{Critical visibilities.} The curves show the parameters $p$ for which the entanglement of the isotropic states in $d\times d$ dimensions become undetectable using the bound of Ref.~\cite{ErkerKrennHuber2017} (upper orange curve) and the bound of~(\ref{eq:quitter bound for I}) for $M=1,2,3$ (blue solid, dashed, dotted curves), respectively. The bottom purple curves indicates the value below which $\rho_{\mathrm{iso}}$ is separable. The irregular behaviour of the curves for $M>1$ originates from the fact that the bases we use are all unbiased w.r.t. each other only in prime dimensions (green dots).}
\label{fig:critical vis isotropic state}
\end{figure}

\noindent
where we have inserted the fidelity bound $\tilde{F}\suptiny{0}{0}{(M)}$ for multiple MUBs derived in Sec.~\ref{sec:appendix more than one tilted basis}. The measurements performed to lower-bound the entanglement dimensionality of $\rho$ may hence directly be used to also obtain a lower bound on the entanglement of formation.

We further note that the bound of (\ref{eq:quitter bound for I}) can also be considered to be a generalization of the bounds discussed in Ref.~\cite{ErkerKrennHuber2017}, where a similar, but strictly weaker bound for $I(\rho)$ is provided, corresponding to setting $M=1$ and $\tilde{\gamma}_{mm\pr nn\pr}\suptiny{0}{0}{(M)}\rightarrow1$. To provide direct comparisons of our bounds with the methods of Ref.~\cite{ErkerKrennHuber2017}, we again turn to the example of the isotropic state $\rho_{\hspace*{0.5pt}\mathrm{iso}}= p\ket{\hspace*{-0.5pt}\Phi^{+}\!}\!\!\bra{\!\Phi^{+}\!}+\tfrac{1-p}{d^{2}}\mathds{1}$, where $0\leq p\leq1$, $\ket{\hspace*{-0.5pt}\Phi^{+}\!}=\tfrac{1}{\sqrt{d}}\sum_{n}\ket{nn}$, and $\mathds{1}$ is the identity in dimension $d^{2}$. A comparison of the performance of these bounds for entanglement quantification for the assumed state $\rho_{\hspace*{0.5pt}\mathrm{iso}}$ is shown in Fig.~\ref{fig:EoF isotropic state}.

The isotropic state also provides an ideal theoretical testing ground for the noise robustness of these bounds, since it corresponds to mixing a maximally entangled state with white noise and hence allows to characterize the robustness of the entanglement bounds against decoherence. To this end, we compare the critical visibilities $p_{\mathrm{crit}}$, that is, the parameters appearing in $\rho_{\mathrm{iso}}(p)$ for which the different methods stop detecting entanglement. Ideally, this could be the case for the value $p_{\mathrm{crit}}=\tfrac{1}{d+1}$, below which the isotropic state is separable~\cite{HorodeckiRPMK2007}. For the bound of Ref.~\cite{ErkerKrennHuber2017} we find $p_{\mathrm{crit}}\suptiny{0}{0}{\mathrm{BW}}=\tfrac{d^{2}-3d+4}{d^{2}-2d+4}$, whereas our bound from~(\ref{eq:quitter bound for I}) provides $p_{\mathrm{crit}}\suptiny{0}{0}{(M)}=\tfrac{d(d-1)+f(M)}{d(d^{2}-1)+f(M)}$, where $f(M)=\sum_{\substack{m\neq m\pr,  m\neq n \\ n\neq n\pr, n\pr\neq m\pr}}\tilde{\gamma}\suptiny{0}{0}{(M)}_{mm\pr nn\pr}$. As illustrated in Fig.~\ref{fig:critical vis isotropic state}, the improved bounds presented here significantly improve on the noise resistance of the bounds.


\subsection{Multipartite entanglement certification}\label{sec:appendix multipartite}

{In this appendix,} we give a brief outlook on the multipartite case. For {this} purpose we define a {family of} generalized GHZ states {for arbitrary local dimension and arbitrary weights $\{\lambda_i\}_{i}$} as
\begin{equation}
    \ket{\mathrm{GHZ}_{\lambda,n,d}}:=\sum_{i=0}^{d-1} \lambda_i \ket{i}^{\otimes n}\,,
\end{equation}
with $\sum_i\lambda_i^2 =1$. {The GHZ-weights $\lambda_i$} can be interpreted as generalized Schmidt coefficients {for this particular family of states} and our fidelity method {can be applied in full analogy to the bipartite cases discussed previously. As, before,} we can introduce {local} tilted bases for the $n$-partite case {as}
\begin{equation}
    \ket{\tilde{j}\suptiny{1}{0}{(n)}}:= \frac{1}{\sqrt{\sum_{k} \lambda^{2/n}_k}}\sum_{m=0}^{d-1} \omega^{jm} \lambda^{1/n}_m \ket{m},
    \label{eq:tilted basis multipartite}
\end{equation}
such that $\ket{\tilde{j}\suptiny{0}{0}{(n=2)}}\equiv\ket{\tilde{j}}$ coincides with our previous definition for bipartite systems. We are then interested in determining a fidelity bound $\tilde{F} (\rho, \mathrm{GHZ}_{\lambda,n,d})$ such that
%
\begin{align}
    F (\rho, \mathrm{GHZ}_{\lambda,n,d}) &:=\text{Tr}(\rho\ket{\mathrm{GHZ}_{\lambda,n,d}}\bra{\mathrm{GHZ}_{\lambda,n,d}})\nonumber\\
    &\geq \tilde{F} (\rho, \mathrm{GHZ}_{\lambda,n,d})\,.
    \label{eq:multipartite Fid}
\end{align}
{Such a bound can indeed be found and, as we shall see, it takes the form}
\begin{align}
    \tilde{F} (\rho, \mathrm{GHZ}_{\lambda,n,d})    &:= \Bigl( \sum_{k} \lambda^{2/n}_k \Bigr)^{n} \bra{\tilde{0}\suptiny{1}{0}{(n)}}^{\otimes n} \hspace*{-1.5pt}\rho \ket{\tilde{0}\suptiny{1}{0}{(n)}}^{\otimes n}
    \label{eq:fid bound multipartite appendix}\\
    &\ \ -\sum_{(\alpha,\beta)\in\gamma}\lambda_{\alpha} \lambda_{\beta}\sqrt{\bra{\alpha}\rho\ket{\alpha}\bra{\beta}\rho\ket{\beta}}.\nonumber
\end{align}
{where $\alpha=(i_1, \dots, i_n)$ and $\beta=(j_1, \dots, j_{n})$ are multi-indices with $i_{k},j_{l}\in\{0,1,\dots,d-1\}$, and we have used the notation $\ket{\alpha}=\ket{i_1, \dots, i_n}$ and
$\lambda_{\alpha}:= \prod_{i_k \in \alpha} \lambda^{1/n}_{i_{k}}$. The sum in the second line of Eq.}~(\ref{eq:fid bound multipartite appendix}){ runs over pairs of multi-indices in the set $\gamma$, which is given by}
\begin{align}
    \gamma  &:= \{ (\alpha, \beta) | \alpha\notin \gamma_{\alpha}\vee\beta\notin\gamma_{\beta}\},
\end{align}
and $\gamma_{\alpha}:=\{\alpha=(i,i,\dots,i)|i=0,1,\dots,d-1\}$ are the sets of multi-indices where all sub-indices $i_{k}$ are the same.

To prove {the relation of Eq.}~(\ref{eq:fid bound multipartite appendix}), we expand the all-zero diagonal element in the tilted basis {w.r.t. the standard basis, that is, inserting from Eq.}~(\ref{eq:tilted basis multipartite}){ we write}
\begin{equation}
    \bra{\tilde{0}\suptiny{1}{0}{(n)}}^{\otimes n} \rho \ket{\tilde{0}\suptiny{1}{0}{(n)}}^{\otimes n} =
    \Bigl( \sum_{k} \lambda^{2/n}_k \Bigr)^{-n} \sum_{\alpha,\beta} \lambda_{\alpha} \lambda_{\beta} \bra{\alpha}\rho\ket{\beta}
\end{equation}
and observe that, just as in the bipartite case, all density matrix elements appear. We can {then use this} to replace terms {in the fidelity} on the left-hand side of {Eq.}~(\ref{eq:multipartite Fid}){ i.e.,}
\begin{align}
    &F (\rho, \mathrm{GHZ}_{\lambda,n,d}) = \sum_{i,j} \lambda_{i} \lambda_{j} \bra{i}^{\otimes n} \hspace*{-1.5pt}\rho \ket{j}^{\otimes n}
    \label{eq:multi fid proof}\\
    &\ \ = \left( \sum_{k} \lambda^{2/n}_{k} \right)^{n} \bra{\tilde{0}\suptiny{1}{0}{(n)}}^{\otimes n} \hspace*{-1.5pt}\rho \ket{\tilde{0}\suptiny{1}{0}{(n)}}^{\otimes n}
    - \sum_{(\alpha,\beta)\in\gamma}\lambda_{\alpha} \lambda_{\beta} \bra{\alpha}\rho\ket{\beta} \,.
    \nonumber
\end{align}
Now{,} we invoke the Cauchy-Schwarz inequality $|\bra{\alpha}\rho\ket{\beta}|\leq\sqrt{\bra{\alpha}\rho\ket{\alpha}\bra{\beta}\rho\ket{\beta}}$ {to bound the last term in Eq.}~(\ref{eq:multi fid proof}){ as we have done in the case of bipartite states, such that we get
\begin{align}
    F (\rho, \mathrm{GHZ}_{\lambda,n,d}) &\geq \Bigl( \sum_{k} \lambda^{2/n}_k \Bigr)^{n} \bra{\tilde{0}\suptiny{1}{0}{(n)}}^{\otimes n} \hspace*{-1.5pt}\rho \ket{\tilde{0}\suptiny{1}{0}{(n)}}^{\otimes n}\nonumber\\
    &\ -
    \sum_{(\alpha,\beta)\in\gamma}\lambda_{\alpha} \lambda_{\beta}\sqrt{\bra{\alpha}\rho\ket{\alpha}\bra{\beta}\rho\ket{\beta}}
    \nonumber\\
    &=\tilde{F} (\rho, \mathrm{GHZ}_{\lambda,n,d})\,.
\end{align}
Note that in the case that {$\rho = \ket{\mathrm{GHZ}_{\lambda,n,d}}\!\!\bra{\mathrm{GHZ}_{\lambda,n,d}}$} all the elements in the sum over {$(\alpha,\beta)\in\gamma$} vanish, as only terms $\bra{i}^{\otimes n} \hspace*{-1.5pt}\rho \ket{j}^{\otimes n}$ {appear} and (\ref{eq:multipartite Fid}) becomes an equality. {This} shows that it is in principle possible to certify a {unit} fidelity with a multipartite and multi-dimensional target state for any $n$ and $d${. However,} using only a single {tilted basis element $\ket{\tilde{0}\suptiny{1}{0}{(n)}}$ comes at the expense of reduced noise resistance, as we have seen in the bipartite case. Although this leaves room for improving the bound by the inclusion of additional tilted basis elements, the practical optimization over all potential combinations of phases is beyond the scope of this brief outlook.}


\subsection{Effects of a wrong choice of Schmidt basis on the fidelity lower bounds}\label{sec:wrong schmidt basis}

{In this section we provide an example of how a choice of standard basis that does not correspond exactly to the Schmidt basis of the generated state affects the value of our fidelity lower bound $\tilde{F}(\rho,\Phi)$. Our example is based on the physically motivated situation in which there is a misalignment between the local reference frames of each party.

For the two-qutrit maximally entangled state $\ket{\Phi^+_3}=\frac{1}{\sqrt{3}}(\ket{00}+\ket{11}+\ket{22})$ we can assume without loss of generality that one side, Alice, performs the first measurement in the correct Schmidt basis while the other side, Bob, measures in a basis that is rotated w.r.t to Alice's measurement basis. This is due to the $U\otimes U^*$ invariance of the isotropic states. Hence, let Alice measure in the standard basis $\{\ket{0},\ket{1},\ket{2}\}$ and let Bob measure in a one-parameter rotation of a two-dimensional subspace of Alice's basis, namely,}

\begin{align}
\ket{\bar{0}}&=\cos{\theta}\ket{0}+\sin{\theta}\ket{1}\\
\ket{\bar{1}}&=\sin{\theta}\ket{0}-\cos{\theta}\ket{1}\\
\ket{\bar{2}}&=\ket{2}.
\end{align}

{From the results of the measurements in the global product basis $\{\ket{m\bar{n}}\}_{m,n}$, one can compute the target state and the tilted basis for each party and complete the procedure outlined in Fig.}~\ref{fig:Adaptive strategy for certifying entanglement} {of the main text to obtain a fidelity lower bound and a certified Schmidt number. The results for this case are plotted in Fig.}~\ref{fig:wrong schmidt basis} {for this example.

This result illustrates how a sub-optimal choice of Schmidt basis can lead to suboptimal fidelity bounds and certified entanglement dimensionality. Crucially, however, it does not invalidate our method as the certified fidelity and entanglement are nonetheless valid. Moreover, one can see that, in our example, small deviations do not cause our fidelity bound to drop drastically, on the contrary, one can still certify maximal entanglement dimensionality up to at least $20\%$ rotation.}

\begin{figure}[t!]
	\includegraphics[width=0.45\textwidth,trim={0cm 0mm 0cm 0mm}]{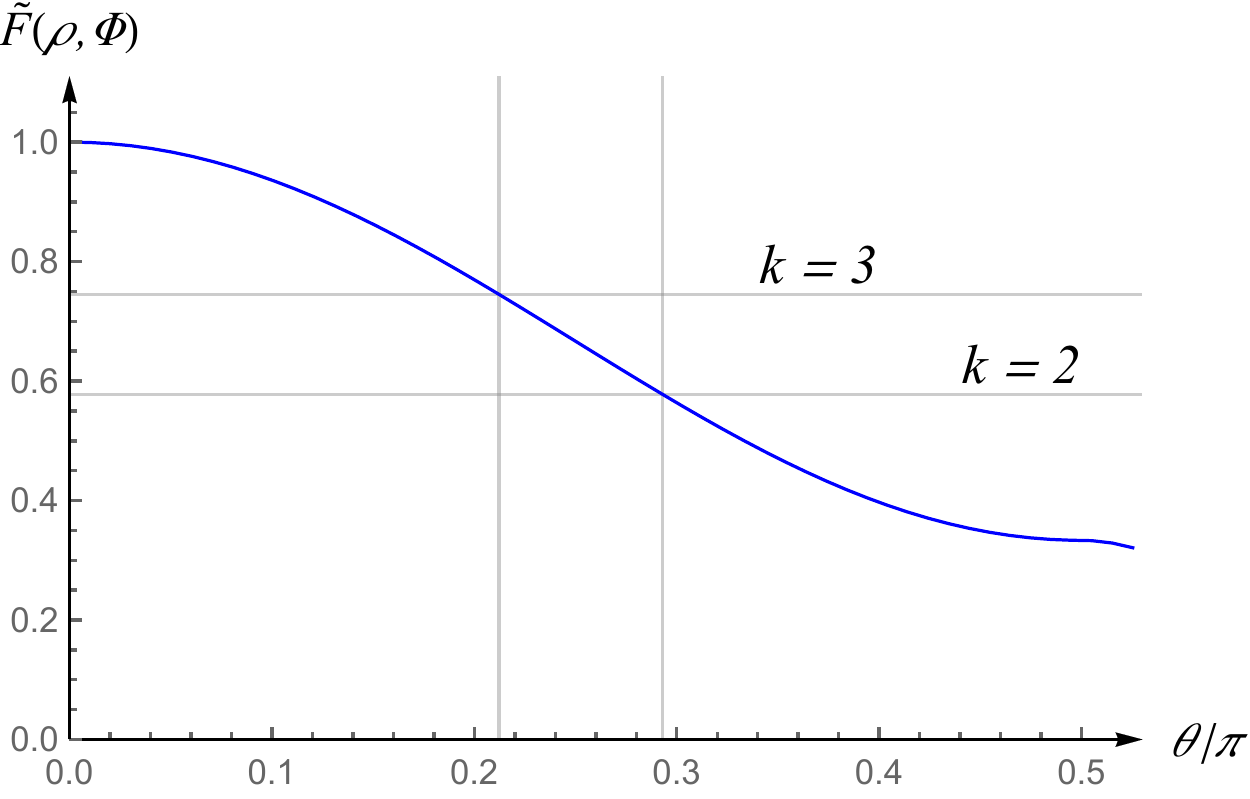}
\vspace*{-2mm}
\caption{Plot of fidelity lower bound $\tilde{F}(\Phi^+_3,\Phi)$ for the maximally entangled two-qutrit state as a function of the rotation angle $\theta$ when one of the sides measures in a standard basis that is rotated in a two-dimensional subspace w.r.t. the measurement basis on the other side. The horizontal lines show the threshold of the fidelity bound in which Schmidt numbers $k=3$ and $k=2$ can be certified.}
\label{fig:wrong schmidt basis}
\end{figure}




\subsection{Classical prepare-and-measure experiment: LG basis}\label{sec:appendix LG Basis}

Here we demonstrate a classical experiment in which we prepare modes in the standard Laguerre-Gaussian (LG) basis and the first mutually unbiased basis (MUB), and then perform measurements in both bases using the technique discussed in the Methods section. The purpose of this experiment is to perform an unfolded, classical version of our two-photon entanglement setup. Also referred to as the Klyshko advanced-wave picture~\cite{Klyshko:1988eea}, this is equivalent to replacing the crystal with a mirror, propagating light back through one of the detectors, reflecting it at the crystal plane, and then propagating it back to the other detector (compare Fig.~\ref{fig:SuppdataLG}~(b) with the setup figure in the main text). In this manner, we can show that we are able to generate and measure arbitrary complex amplitudes, and that our measured bases are indeed mutually unbiased with respect to each other.

As shown in Fig.~\ref{fig:SuppdataLG}~(b), modes in seven-dimensional LG and MUB bases are generated using computer generated holograms (CGH) implemented on the SLM labelled (g). Intensity images of these modes obtained on a CCD camera are shown in Fig.~\ref{fig:SuppdataLG}~(a). The modes generated by SLM (g) are imaged onto SLM (m) by a 4f system of lenses (l3, 400mm) through a pinhole to pick off the first diffraction order of the SLM and remove zero-order diffraction noise. The pinhole is also where the crystal plane would be in the Klyshko picture (dotted rectangle). A measurement of a particular mode is performed by the spatial-mode filter implemented by SLM (m), a single-mode fiber (SMF), and a single-photon avalanche photodiode (SPAD). The measurement holograms on SLM (m) are scanned through modes in both LG and MUB bases to obtain a 14$\times$14 element matrix of counts shown in Fig.~\ref{fig:SuppdataLG}~(c). The counts are normalised such that the total counts measured across one basis are equal for each generated mode. As can be clearly seen, when modes in the same basis are generated and measured, a strong diagonal matrix is obtained, with a visibility of 94.8\% (LG) and 84.4\% (MUB) \textemdash\ defined as the sum of diagonal counts divided by total counts. The visibility in the LG basis is lower than the near-unity theoretical value due to imperfect alignment. The MUB visibility is further lowered due to errors introduced by the CGH method for approximating a more complex scalar field with a phase-only hologram, which is confirmed by simulation. When the generation and measurement bases are different, the data sets are seen to be mutually unbiased (flat), with a visibility approaching 1/7=14.3\% in both cases (15.6\% and 13.5\%).

\begin{figure*}
	\includegraphics[width=0.65\textwidth,trim={0cm 0mm 0cm 0mm}]{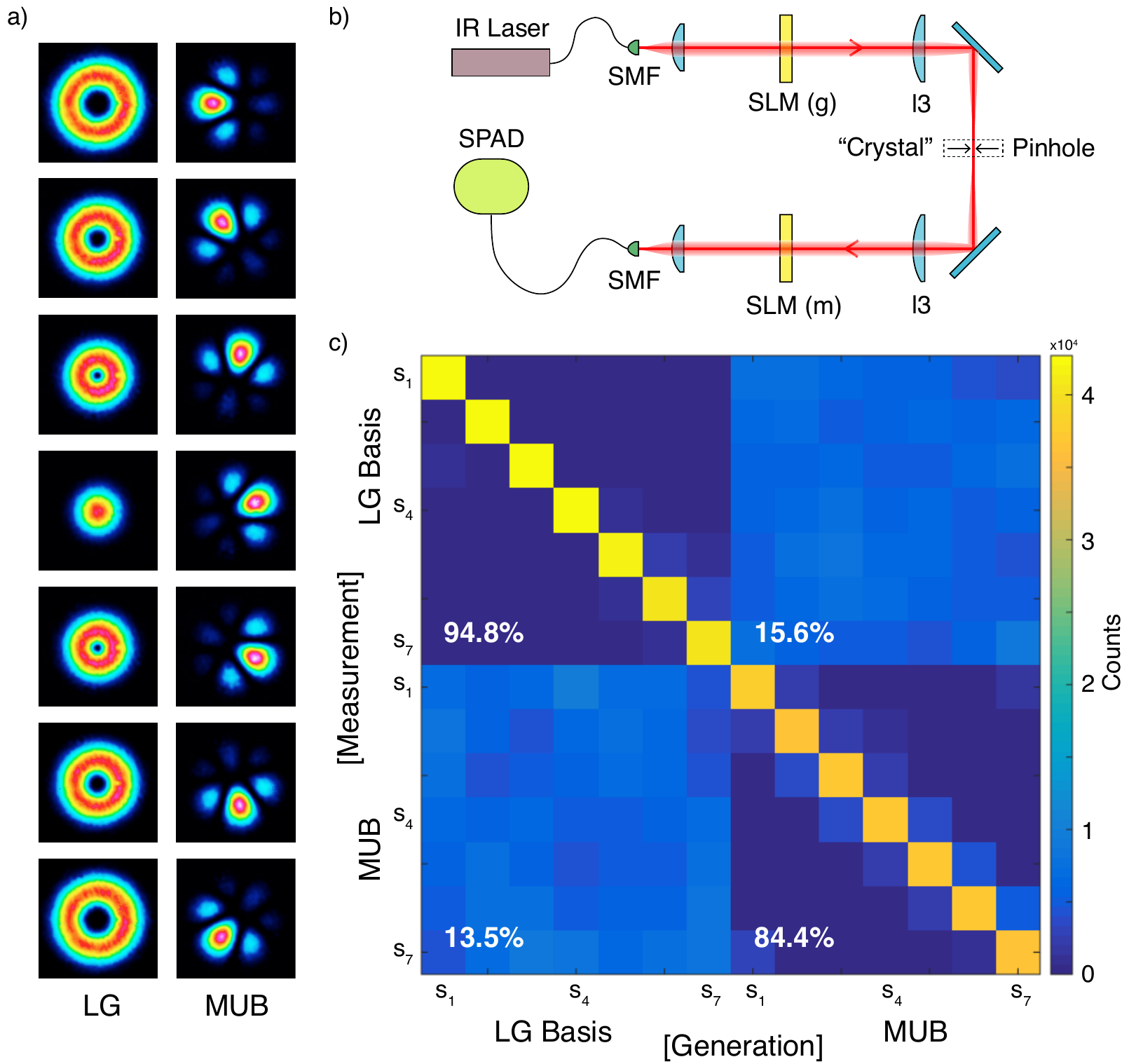}
\vspace*{-2mm}
\caption{\textbf{Classical prepare-and-measure experiment: LG basis.} a) CCD images of the 7-dimensional Laguerre-Gaussian (LG) basis and first mutually unbiased basis (MUB) modes. b) The experiment consists of a strongly attenuated IR laser incident on a spatial light modulator (SLM (g)) used for generating arbitrary spatial modes. SLM (g) is imaged onto SLM (m), which displays measurement holograms for arbitrary spatial modes. A pinhole is used to remove zero-order diffraction noise from SLM (g), and is also located at the ``crystal" plane in the unfolded Klyshko picture \cite{Klyshko:1988eea}. The light from SLM (m) is coupled into a single-mode fiber (SMF), which is connected to a single-photon avalanche diode (SPAD). c) Experimental data showing measured counts when states are prepared and measured in both bases. The data are strongly correlated when the preparation and measurement bases are the same, and completely uncorrelated when they are not.}
\label{fig:SuppdataLG}
\end{figure*}

\begin{figure*}
	\includegraphics[width=0.7\textwidth,trim={0cm 0mm 0cm 0mm}]{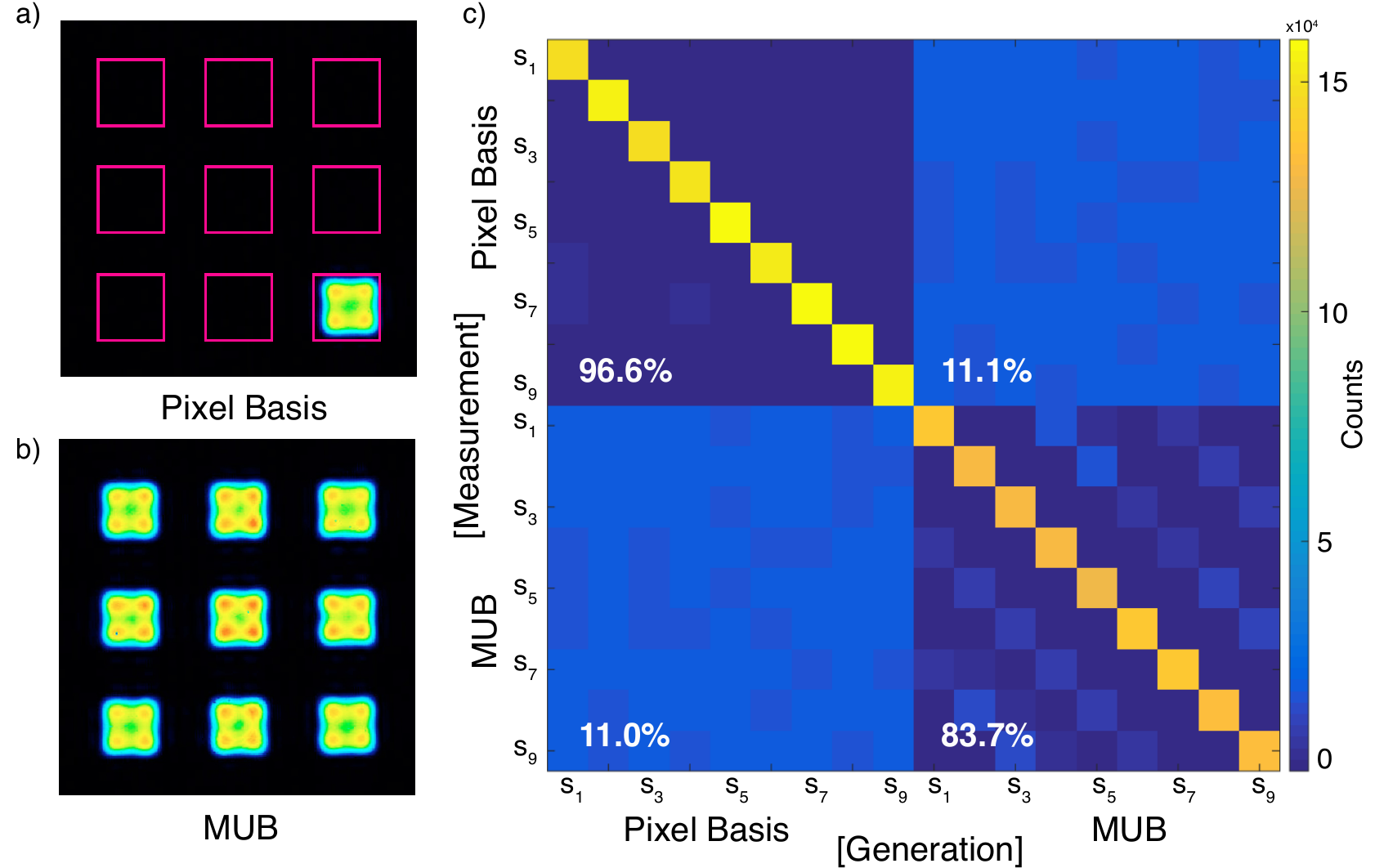}
\vspace*{-2mm}
\caption{\textbf{Classical prepare-and-measure experiment: 9-dimensional pixel basis.} CCD images of a) the first Pixel basis mode, and b) first mutually unbiased basis (MUB) mode. c) Experimental data showing measured counts when states are prepared and measured in both bases. The data are strongly correlated when the preparation and measurement bases are the same, and completely uncorrelated when they are not.}
\label{fig:SuppdataPX}
\end{figure*}

\subsection{Examples of MUBs in other experimental degrees-of-freedom/platforms}\label{sec:appendix Implementations of MUBs in other setups}

The purpose of this section is to show that our entanglement certification technique can be readily applied to other photonic degrees-of-freedom (DOFs), as well as to other physical platforms such as atoms. We do this by first demonstrating a second set of measurement bases with our classical prepare-and-measure experiment: the Pixel basis. Then, we discuss recent experimental examples of mutually unbiased measurements in the time-frequency and path degrees-of-freedom. Finally, building on recent experiment results, we show how such measurements are also feasible in high-dimensional atomic systems consisting of Cesium atoms.

First, we use the classical prepare-and-measure experiment discussed in Sec.~\ref{sec:appendix LG Basis} to demonstrate a second set of mutually unbiased bases for the photonic position-momentum DOF. As shown in Fig.~\ref{fig:SuppdataPX}~(a), the Pixel basis is composed of nine position states, defined by nine discrete macro-pixels. The figure shows the intensity profile of the first Pixel basis state recorded on a CCD, with the eight empty boxes indicating the positions of the remaining Pixel basis states. Fig.~\ref{fig:SuppdataPX}~(b) shows the intensity profile of the first state from the first mutually unbiased basis (MUB) to the Pixel basis, constructed according to the standard method discussed in Ref.~\cite{WoottersFields1989}. Using the setup from Fig.~\ref{fig:SuppdataLG}~(b), every state in the Pixel and MUB bases is generated using SLM (g) and imaged onto the measurement SLM (m). The measurement SLM (m) is used to display measurement holograms in both bases, resulting in an 18$\times$18 element matrix of counts shown in Fig.~\ref{fig:SuppdataPX}~(c). As can be clearly seen, when modes in the same basis are generated and measured, a strong diagonal matrix is obtained, with a visibility of 96.6\% (Pixel) and 83.7\% (MUB). As for the LG basis, the MUB visibility is slightly lower than the Pixel basis due to errors introduced by the CGH. When the generation and measurement bases are different, the data are again seen to be mutually unbiased (flat), with a visibility approaching 1/9=11.1\% in both cases (11.1\% and 11.0\%).

Despite the significant difficulties in the implementation of arbitrary measurements on high-dimensional quantum systems, measurements in specific bases (such as MUBs) are quite common, with recent advances allowing for this in several experimental platforms. Here we briefly discuss how mutually unbiased and tilted basis measurements can be implemented in these platforms, allowing our entanglement certification technique to be directly applied in a wide range of future experiments. While we have demonstrated precise control and measurement over photonic spatial modes, recent experiments have been performed that show similar capabilities for other high-dimensional DOFs such as time-frequency and path.

For example, the experiment of Kues et al.~\cite{Kues:2017db} demonstrates on-chip, high-dimensional frequency-mode entanglement via spontaneous four-wave mixing in a micro-ring resonator. In order to measure their entangled state, the authors use a combination of two programmable phase filters and an electro-optic phase modulator to perform projective measurements corresponding to state vectors of the form $\ket{\psi_{\textrm{proj}}} = \sum_{k=0}^{d-1} \alpha_k e^{i\phi_k}\ket{\bar{k}+k}$ where the projection amplitudes $\alpha_k$ and the phases $\phi_k$ can be chosen arbitrarily for a given frequency mode $\bar{k}$. This is precisely the type of transformation that would be required for a measurement in an arbitrary tilted or mutually unbiased basis of frequency modes, allowing the direct application of our method to this platform.

Another recent experiment by Karpi\'{n}ski et al.~\cite{Karpinski:2016bm} used an electro-optic modulator to carry out a temporal Fourier transform of heralded single-photon pulses, while preserving their quantum coherence. This ``time lens" applies the exact transformation required to measure temporal pulse-mode-entangled states in the mutually unbiased frequency basis. In the recent experiment by Carolan et al.~\cite{Carolan:2015fb}, the authors demonstrate exquisite control over a rapidly reprogrammable 6-mode integrated photonic circuit, implementing Haar random unitaries with an extremely high fidelity. Combined with multi-outcome measurements at the end of the linear circuit, their system can readily be used to perform measurements in a six-dimensional mutually unbiased basis of path modes.

In the recent experiment by Anderson et al.~\cite{Anderson:2015fu}, the electron and nuclear spins of individual ${}^{133}$Cs atoms were used as a test bed for implementing high-dimensional unitary transformations on an atomic system. Radio frequency and microwave magnetic fields were used to generate control Hamiltonians with excellent performance even in the presence of static and dynamic perturbations, allowing the implementation of unitary maps in a 16-dimensional Hilbert space with fidelities greater than 0.98. This was followed by a Stern-Gerlach measurement apparatus that measured the population in the 16-dimensional Hilbert space. Together, these unitary operations and multi-outcome measurements are precisely what is required to measure in a mutually unbiased basis of electron and nuclear ${}^{133}$Cs atoms spins.

Finally, one may note that multi-qubit systems, such as have been realized in photonic systems~\cite{ValloneEtAl2009, GaoEtAl2010, WangEtAl2016, WangEtAl2018}, superconducting qubits~\cite{SongEtAl2017}, or trapped ions~\cite{FriisEtAl2017}, can also yield subsystems with high local dimension for suitable bipartitions of groups of multiple qubits. Such platforms are often composed of individually controllable qubits, e.g., for quantum computation or simulation~\cite{FriisEtAl2017}, and usually permit arbitrary (projective) single-qubit measurements, and hence allow measurements w.r.t. mutually unbiased or tilted bases for any bipartition. For instance, measurements w.r.t. the local Pauli $Z$ and $X$ operators for all qubits would be a simple realization of a MUB measurement. Our methods are thus also applicable to such systems.

The above examples clearly demonstrate the wide applicability of our entanglement certification technique to a variety of physical systems, and highlights its potential for significantly impacting future experiments on high-dimensional entanglement in photonic and atomic platforms, and beyond.


\subsection{Systematic errors}\label{sec:syserror}

While there are no assumptions made about the state or how it is produced, the method introduced here intrinsically puts trust on the measurement devices to work correctly. Hence, a crucial part of the experiment is an in-depth characterization of the measurement method. While from a physical point of view, one would expect the crystal to predominantly produce perfectly correlated pairs due to a conservation of angular momentum, the real data features a significant amount of cross-talk and noise, ultimately diminishing the certified entanglement and dimensionality. On the other hand, non-perfect unbiasedness of the observables could even lead to classically correlated photons to appear entangled, the most extreme case being a measurement in two identical bases that while assumed to be unbiased, are actually the same. Furthermore, the coincidence counts in different settings may not correspond to the density matrix elements in the way assumed if the detector efficiency is different for the different bases, which could lead to either over- or under-estimation of correlations (and with it entanglement). These are all potential systematic errors that we want to address in this final section.

\begin{smallboxtable}{Systematic error due to \\ imperfect measurements}{syserror2}
\begin{center}
{\renewcommand{\arraystretch}{1.5}
\begin{tabularx}{\textwidth}{|Y||Y|Y|Y|}
	\hline
	\ $d$ \   & \ $\tilde{F}(\rho,\Phi^{+})$\  & \ $\tilde{F}_{\textrm{s1}}(\rho,\Phi^{+})$\  & \ $\tilde{F}_{\textrm{s2}}(\rho,\Phi^{+})$\ \\
	\hline
	3          & 91.5$\pm$0.4\%  & 98.0\% & 95.6\% \\
	5          &  89.9$\pm$0.4\% &  96.4\% & 92.3\% \\
	7          &  84.2$\pm$0.5\% & 94.6\%  & 87.6\% \\
	11        &  74.8$\pm$0.4\% & 89.7\% & 80.6\% \\
	\hline
\end{tabularx}\\
}
\end{center}
\small{$\tilde{F}(\rho,\Phi^{+})$ and $\tilde{F}_{\textrm{s1/2}}(\rho,\Phi^{+})$ are experimental and simulated fidelities to the maximally entangled state obtained via measurements in two MUBs in dimension $d$, respectively. $\tilde{F}_{\textrm{s1}}(\rho,\Phi^{+})$ is obtained by incorporating the effects of imperfect hologram measurements on an ideal state estimated from measurements in the LG basis. $\tilde{F}_{\textrm{s2}}(\rho,\Phi^{+})$ is obtained by additionally taking into account the misalignment-induced crosstalk measured in the LG basis.
}
\end{smallboxtable}

While the predominant source of crosstalk is due to imperfect alignment, our paradigm of state-independence also includes the notion of reference frames (i.e.~we do not assume to have a perfect common reference frame) and this misalignment can only decrease observed correlations. In other words, alignment issues are essentially captured by local unitary operations and can never lead to an increase of correlations where there are none.

Upon inspecting the correspondence of coincidence counts to density matrix elements we noticed a significant impact of mode-dependent loss. The usual assumption that coincidence counts $C_{ij}$ of $N$ photon pairs per unit of time in basis elements $i$ and $j$ respectively are related to density matrix elements via
\begin{align}
C_{ij}=N\langle ij|\rho|ij\rangle \,,
\end{align}
implicitly assumes (1) a constant photon pair production rate and (2) a universal coupling efficiency that is independent of $i$ and $j$. While the measured pair production rate fluctuations are low enough for that estimation to be valid, we actually do expect a strong mode-dependent loss. In the LG-basis we expect from theoretical computations that higher modes have a lower coupling efficiency in the single-mode fibers \cite{QassimEtAl2014}, which should lead to a systematic suppression of higher-mode coincidence counts and with it a systematic under-estimation of entanglement. The exact coupling efficiency, however, depends on many intricate details of the physical setup and any theoretical computation could increase systematic errors in unpredictable ways. In this section we thus introduce a general method that corrects for mode-dependent loss using only the singles and coincidences in the setup and will find application also in many other quantum optical setups. Denoting the singles per unit time in detector $A/B$ as $S_i^{A/B}$, as well as the mode dependent loss factors as $\eta_i^{A/B}$, we conclude that:
\begin{align}
C_{ij}=N\langle ij|\rho|ij\rangle \eta_i^{A}\eta_i^{B}
\end{align}
{as well as}
\begin{align}
S_{i}^{A/B}=N\langle i|\rho_{A/B}|i\rangle \eta_i^{A/B}
\end{align}
{Now if we define}
\begin{align}
M_{ij}:=\frac{C_{ij}}{S_{i}^AS_{j}^B}=\frac{1}{N}\frac{\langle ij|\rho|ij\rangle}{\langle i|\rho_A|i\rangle\langle j|\rho_B|j\rangle}
\end{align}
{we can use the fact that}
\begin{align}
\sum_j M_{ij}\langle j|\rho_B|j\rangle=[M\vec{\rho_B}]_i=\frac{1}{N}\frac{\sum_j\langle ij|\rho|ij\rangle}{\langle i|\rho_A|i\rangle}=\frac{1}{N}
\end{align}
{This allows us to get $N$, as well as conclude that}
\begin{align}
\langle i|\rho_B|i\rangle=\sum_j (M)^{-1}_{ij}\frac{1}{N}
\end{align}
{Now all that is left is to insert this into the definition of $M_{ij}$ to get}
\begin{align}
\langle ij|\rho|ij\rangle=\frac{M_{ij}(\sum_j (M)^{-1}_{ij})(\sum_i (M^T)^{-1}_{ij})}{(\sum_i\sum_j (M)^{-1}_{ij})}
\end{align}

The only assumptions remaining in this correction method are a constant pair production rate and that the majority of singles is generated by photon pairs. These assumptions can also be verified using the experimental data by checking that $[M\vec{\rho_B}]_i^{-1}=N$ is indeed equally true for all $i$. Using this correction method we indeed find the expected effect: higher order modes in LG basis were significantly suppressed leading to artificially reduced coincidence counts. We account for this mode-dependent loss in our data, allowing us to more accurately estimate a target state and hence construct a more optimum tilted basis.

A second source of systematic error is the effect of imperfect measurements on the resulting fidelity bounds. As shown in Sec.~\ref{sec:appendix LG Basis}, the classical (one-photon) measurements made using our computer-generated holograms (CGHs) in the LG and the MUB bases are not perfect, with the MUB basis showing a lower visibility than the LG basis. In the two-photon experiment, this would manifest as additional counts appearing in the off-diagonals of the data matrices shown in the main text, {which would in turn lower the measured fidelity bounds. In order to estimate this quantitatively, we proceed as follows.

First, we calculate the ideal state as obtained from diagonal measurements in the LG basis}, {by setting the off-diagonal (crosstalk) counts to zero and calculating the resulting density matrix. Second, we use this state to calculate the ideal experimental data one would obtain if measuring in the first MUB}. Next, we simulate the imperfect measurements in MATLAB for each input state and hologram by multiplying the complex field amplitude by the hologram amplitude, and then calculating its overlap with a Gaussian fiber mode amplitude. The resulting probability matrices for the LG and MUB bases capture the resulting imperfections of the CGH measurement process. This process is repeated for each dimension considered in our experiment. We find that the imperfections in the LG measurement are almost negligible, while the visibility in the MUB drops as a function of dimension. We then calculate the effect of these hologram imperfections on the ideal two-photon experimental data calculated above.

A key factor that results in a lowering of the measured fidelity bound in experiment is the crosstalk due to imperfect alignment. We incorporate this crosstalk into our fidelity calculation by using the LG basis data obtained in experiment, and the MUB data obtained from the above simulation. In this manner, both the effects of imperfect measurement and misalignments are captured in our systematic error-corrected fidelity bounds. Table~\ref{table:syserror2} lists the measured fidelities $\tilde{F}(\rho,\Phi^{+})$ from experiment, the simulated fidelities $\tilde{F}_{\textrm{s1}}(\rho,\Phi^{+})$ taking into account the effects of imperfect hologram measurements, and simulated fidelities $\tilde{F}_{\textrm{s2}}(\rho,\Phi^{+})$ additionally incorporating the effects of misalignment-induced crosstalk only in the LG basis (taken from the measured data). The effects of crosstalk on the MUB measurements cannot be added in independently of the simulated systematic error, but one can expect that it will lower the fidelities even more, ideally approaching the measured values $\tilde{F}(\rho,\Phi^{+})$. Thus, imperfect measurements are always seen to result in an under-estimation of correlations, thus lowering our fidelities from their ideal expected values.


\bibliographystyle{mod_unsrt}
\bibliography{biblio_suppl}